\documentclass[]{iopart}

\expandafter\let\csname equation*\endcsname\relax
\expandafter\let\csname endequation*\endcsname\relax

\usepackage{amsfonts}
\usepackage{amsmath}
\usepackage{amssymb}
\usepackage{bm}
\usepackage{dcolumn}
\usepackage{graphicx}
\usepackage{graphics}
\usepackage[T1]{fontenc}
\usepackage[utf8]{inputenc}
\usepackage{latexsym}
\usepackage{rotating}   
\usepackage[colorlinks=true]{hyperref}
\usepackage[all]{hypcap}
\usepackage{xspace}
\usepackage[usenames]{color}
\usepackage{microtype}

\usepackage{ulem}
\definecolor{darkgreen}{rgb}{0.2,0.7,0.2}

\newcommand\bw{}
\newcommand\ew{}

\newcommand{\nn}{\nonumber}
\newcommand{\pont}{{^\ast\!}RR}

\newcommand{\pd}{\partial}
\newcommand{\cd}{\nabla}

\newcommand{\gr}{{\mbox{\tiny GR}}}

\newcommand{\eh}{{\mbox{\tiny EH}}}
\newcommand{\cs}{{\mbox{\tiny CS}}}
\newcommand{\ay}{{\mbox{\tiny AY}}}
\newcommand{\Mat}{{\mbox{\tiny Mat}}}

\newcommand{\MS}[1]{\textrm{\tiny $#1$}}
\newcommand{\ZO}{\MS{(0)}}
\newcommand{\FO}{\MS{(1/2)}}
\newcommand{\SO}{\MS{(1)}}

\newcommand{\FOZO}{\MS{(1/2,0)}}

\newcommand{\FOSO}{\MS{(1/2,2)}}

\newcommand{\FOFFO}{\MS{(1/2,4)}}
\newcommand{\SOZO}{\MS{(1,0)}}

\newcommand{\notu}{\psi}

\newcommand{\TR}{\tilde{r}}

\newcommand{\tD}{\tilde{\Delta}}
\newcommand{\tS}{\tilde{\Sigma}}
\newcommand{\tsf}{\tilde{\vartheta}}
\newcommand{\tg}{{g}}

\newcommand{\dub}{\,\,}
\newcommand\nts{\negthickspace}

\newcommand{\eps}{\varepsilon}

\DeclareMathOperator{\arccot}{arccot}

\begin{document}
\title{Extremal Black Holes in Dynamical Chern-Simons Gravity}

\author{Robert McNees}
\address{Loyola University Chicago, Department of Physics, Chicago, IL 60660, USA.}
\address{Kavli Institute for Theoretical Physics, University of California, Santa Barbara, CA 93106, USA.}

\author{Leo C.~Stein}
\address{TAPIR,
Walter Burke Institute for Theoretical Physics,
California Institute of Technology, Pasadena, CA
91125, USA}

\author{Nicol\'as Yunes}
\address{Department of Physics, Montana State University, Bozeman, MT 59717, USA.}
\address{Kavli Institute for Theoretical Physics, University of California, Santa Barbara, CA 93106, USA.}
\date{\today}

\begin{abstract}
Rapidly rotating black hole solutions in theories beyond
general relativity play a key
role in experimental gravity, as they allow us to compute
observables in extreme spacetimes that deviate from the predictions of
general relativity.
Such solutions are often difficult to find in beyond-general-relativity theories due to
the inclusion of additional fields that couple to the metric
non-linearly and non-minimally.
In this paper, we consider rotating black hole solutions in one such
theory, dynamical Chern-Simons gravity, where the Einstein-Hilbert
action is modified by the introduction of a dynamical scalar field
that couples to the metric through the Pontryagin density.
We treat dynamical Chern-Simons gravity as an effective field theory
and work in the decoupling limit, where
corrections are treated as small perturbations from general
relativity.
We perturb about the maximally-rotating Kerr solution, the
so-called extremal limit, and develop mathematical insight into the
analysis techniques needed to construct solutions for generic spin.
First we find closed-form, analytic expressions for the extremal
scalar field, and then determine the trace of the metric perturbation, giving both
in terms of Legendre decompositions.
Retaining only the first three and four modes in the Legendre
representation of the scalar field and the trace, respectively, suffices
to ensure a fidelity of over $99\%$ relative to full numerical
solutions.
The leading-order mode in the Legendre expansion of the trace of the
metric perturbation contains a logarithmic divergence at the extremal
Kerr horizon, which is likely to be unimportant as it occurs inside
the perturbed dynamical Chern-Simons horizon.
The techniques employed here should enable the construction of
analytic, closed-form expressions for the scalar field and metric
perturbations on a background with arbitrary
rotation. 
\end{abstract}


\maketitle

\section{Introduction}

Einstein's theory of general relativity (GR) has passed a plethora of Solar
System and binary pulsar tests~\cite{lrr-2014-4}, but it has not 
been tested in depth in the \emph{extreme gravity}
regime~\cite{Yunes:2013dva,Berti:2015itd} where the gravitational interaction is
non-linear and dynamical.
A number of new observations will allow us to test this regime of
Einstein's theory: in the gravitational wave spectrum through Advanced
LIGO and its partners~\cite{Abramovici:1992ah,Giazotto:1988gw}, when compact objects collide; in the radio
spectrum with the Event Horizon Telescope~\cite{Loeb:2013lfa}, when an accretion disk
illuminates its host black hole and creates a `shadow'; and in the
X-ray spectrum with the Chandra Telescope~\cite{Weisskopf:2001uu}, when gas heats up and glows
as it accretes in the black hole spacetime.
Such future observations will either confirm Einstein's theory at
unprecedented levels or reveal new phenomena in the
extreme gravity regime.

Solutions that represent rotating black holes (BHs) in theories of
gravity beyond general relativity are an essential ingredient 
of tests in the extreme gravity regime. Constraining these theories requires a metric
with which to calculate observables. Once a metric is available, 
one can investigate the modal stability of the solution, 
calculate the gravitational waves emitted as
two BHs inspiral, compute the `shadow' cast by a BH when illuminated
by an accretion disk, and determine the energy spectrum of the
radiation emitted by gas accreting into the BH.

One beyond-GR gravity theory in which generic rotating BH solutions
have not yet been found is dynamical Chern-Simons (dCS)
gravity~\cite{Alexander:2009tp}. This theory modifies the
Einstein-Hilbert action by introducing a dynamical (pseudo) scalar
field that couples non-minimally to the metric through the Pontryagin
density. The interaction leads to a scalar field evolution
equation that is sourced by the Pontryagin density, and modified
metric field
equations with third derivatives. The latter have cast doubt on
whether full dCS is well-posed as an initial value
problem~\cite{Delsate:2014hba}, and also on whether stable BH
solutions exist. 
We take the point of view that dCS must be treated as an effective
field theory, since it is motivated from the low-energy limit of
compactified heterotic string theory~\cite{Green:1984sg,Green:1987mn}
(for a review see~\cite{Alexander:2009tp}), from
effective field theories of inflation~\cite{Weinberg:2008hq}, and from
loop quantum gravity~\cite{Taveras:2008yf}. Thus the theory is treated
in the \emph{decoupling limit}: deformations from GR are treated
perturbatively, reducing the order of the field equations.

When treated as an effective theory, BH solutions in dCS have
been found in certain limits. A non-rotating BH was found by
Jackiw and Pi~\cite{Jackiw:2003pm}, who showed that the Schwarzschild metric is also a
solution of dCS. Linear stability of high-frequency waves about the Schwarzschild
background was suggested by~\cite{Ayzenberg:2013wua}.
The first investigations of rotating solutions assumed a small rotation parameter. The axionic hair on a slowly-rotating BH was found in~\cite{Campbell:1990ai}. Later, the metric solution to linear order in spin was found
independently by Yunes and Pretorius~\cite{Yunes:2009hc}, and Konno,
et al.~\cite{Konno:2009kg}.  This solution was extended to
quadratic order by Yagi, Yunes and Tanaka~\cite{Yagi:2012ya}. More recently, Konno and
Takahashi~\cite{Konno:2014qua} and Stein~\cite{Stein:2014xba} investigated the behavior of the dynamical scalar field about a
rapidly rotating Kerr background. Stein also investigated the trace of the metric perturbation, and found that the extremal limit may be singular, which partly motivated the present work.

At present, nobody has succeeded in constructing the full metric of
generic rotating BHs in dCS gravity, despite over two decades of work
in that direction~\cite{Campbell:1990ai, Jackiw:2003pm,
  Grumiller:2007rv, Yunes:2007ss, Sopuerta:2009iy, Yunes:2009hc,
  Konno:2009kg, Yagi:2012ya, Konno:2014qua, Stein:2014xba}. The
general procedure for obtaining the metric in the decoupling limit is
simple enough. First, one finds the leading order behavior of the
scalar field as sourced by the background geometry. The scalar field
determines the source for the leading order term in the deformation of
the metric, which must then be solved for. But progress has been
limited because the initial step has proven difficult; the scalar
field satisfies a complicated partial differential equation on the
Kerr background, and only partial results have been obtained for
finite rotation. Therefore, if the goal is to find a complete
description for the metric deformation, then a full solution for the
scalar field is an unavoidable first step.\footnote{%
  A more modest approach might be to construct a reliable analytical
  approximation for the metric. But this requires understanding the
  scalar field well enough to ensure that the source terms for the
  metric, which depend on the scalar, are rendered with sufficient
  detail. And even if the goal is to develop a purely numerical
  description of the metric, some basic understanding of the scalar is
  still needed to test the robustness of the simulations. Before we
  can attack the problem of the metric deformation, then, we must
  first make more progress on solving the general problem of the
  scalar field.}

For all of these reasons, as a first step toward finding full rotating BH solutions in effective
dCS gravity, we study the scalar field on a background where the BH spin takes the maximal Kerr value.  
This extremal limit is of interest not only because the mathematics
simplify significantly, but also because of the Kerr/CFT
conjecture that posits a dual holographic description in terms of a
two-dimensional conformal field theory~\cite{Guica:2008mu,Ghezelbash:2009gf,Matsuo:2010ut,Carlip:2011ax,Compere:2012jk,Dias:2012pp}. Working in the extremal limit, we obtain a general, closed-form expression for the Legendre
modes of the scalar field. The radial structure of the scalar field
is more complicated than that of the slowly-rotating case.
Whereas the slowly-rotating case only requires a finite polynomial
expansion, the rapid-rotation case is characterized by natural
logarithms and arctangents. And unlike the slowly-rotating case, where the scalar field is primarily dipolar, the octupole mode of the field on the extremal background carries more than half of the field's ADM energy. We find that retaining the first 3 non-vanishing Legendre modes of the scalar field is necessary to achieve a fidelity above
$99\%$ in the entire domain relative to the full numerical solution.
Our results for the scalar field complete the first step in the process outlined above and establish
a viable starting point for either an analytic or numerical treatment of the dCS metric deformation of
the extremal Kerr solution.

With analytic, closed-form expressions for the extremal scalar field
in hand, we turn our attention to the BH metric. Treating the dCS correction
as a small perturbation of the Kerr background and working in a convenient
(Lorenz-like) gauge, we show that the modified field equations for the
trace of the metric perturbation can be solved in quadrature in terms
of another Legendre mode decomposition. As in the scalar field case,
the radial structure of the metric perturbation is quite different
from that of slowly-rotating BHs. In particular, the dominant monopole mode exhibits 
a logarithmic divergence at the extremal Kerr horizon, which confirms a conjecture made
in~\cite{Stein:2014xba} by one of the authors. This divergence, however, may be unphysical 
because the Kerr horizon is likely ``inside'' the perturbed dCS horizon. 
Away from this region, the angular structure of the trace of the metric perturbation is predominantly monopolar, with higher order modes modes playing a more important role than in the slowly-rotating case. We find that retaining the first 4 non-vanishing modes of the trace of the metric perturbation
achieves a fidelity above $99\%$ in the entire domain relative to the full numerical solution. 

Although the trace of the metric perturbation is not a gauge-invariant
observable quantity, and although we have not yet analyzed the full
metric perturbation, the calculation of the trace establishes two
useful results. First, the techniques used to solve the equation of
motion for the scalar field are also applicable to the scalar degree
of freedom in the metric perturbation.  This suggests that similar
techniques can probably also be used to solve for the other degrees of
freedom of the metric perturbation, and work along these lines is
currently underway.  Second, it appears that only a small number of
modes are necessary to accurately approximate both the scalar field
and the metric perturbation.  This suggests that an excellent
approximation to a full numerical solution for the metric perturbation
could be achieved through a spectral decomposition that retains a
finite number of modes.

The results we have obtained also have various other consequences for
the study of BHs in dCS gravity and other beyond-GR gravity theories. First, the techniques used to solve for the scalar field on the
extremal Kerr background can be applied to subsequent terms in a
near-extremal expansion in dCS gravity. More importantly, these
techniques generalize to backgrounds with arbitrary rotation
parameter~\cite{Stein:2014xba}, as well as other higher-curvature
interactions. Some of our results have already been extended to
backgrounds with arbitrary spin, and will be presented
elsewhere~\cite{Bob-and-Leo-in-prep}.
Second, the scalar degree of freedom in the metric perturbation can be studied using the same methods we employed for the scalar field. It seems likely that the vector and transverse-traceless components of the metric perturbation can be analyzed using similar techniques, and that these techniques can also be applied to Einstein-dilaton-Gauss-Bonnet gravity and other quadratic gravity theories~\cite{Kanti:1995vq,Alexeev:1996vs,Torii:1996yi,Kleihaus:2011tg,Yunes:2011we,Pani:2011gy,Ayzenberg:2014aka}.
Third, the logarithmic divergence of the trace of the metric suggests a
conjecture: that the dCS-corrected horizon is ``outside'' of the Kerr
horizon for all possible values of angular momentum, protecting against a
naked singularity. The verification of this conjecture will require further work
that utilizes the solution for the full metric perturbation. 
And finally, our results also suggest that
generic rotating BH solutions in beyond-GR theories will not have the
simple rational-polynomial form that the Kerr metric enjoys, and may
require more complicated functional forms. If true, this would imply
that the simple parameterized metrics used to constrain GR deviations 
with Event Horizon Telescope observations of the BH shadows cannot be
used to constrain quadratic gravity theories.

The remainder of this paper presents details of the techniques
developed, the solutions obtained, and their properties. Henceforth, we
use the following conventions. Latin letters
$(a,b,c,\ldots)$ in index lists stand for spacetime
indices.  Parentheses and
square brackets in index lists stand for symmetrization and
anti-symmetrization respectively. The metric signature will be
$(-,+,+,+)$ and we choose units in which $c=1$. However, we do not set
$G$ or $h$ to unity. All other conventions follow the standard
treatment of~\cite{Misner:1974qy,Wald:1984rg}.

\section{The ABC of dCS}
\label{sec:ABC-DCS}

Dynamical Chern-Simons
gravity~\cite{Jackiw:2003pm,Alexander:2009tp} is a four-dimensional
theory defined by the action
\begin{align}
I = I_{\eh} + I_{\cs} + I_{\vartheta} + I_{\Mat}\,.
\label{eq:full-action}
\end{align}
The first term is the Einstein-Hilbert action
\begin{align}
I_{\eh}  = \int d^{4}x \; \sqrt{-g} \; \left(\frac{1}{2 \kappa^{2}} \, R\right)\,,
\end{align}
where $\kappa^{2} = 8 \pi G$, $R$ is the Ricci scalar associated with
the metric tensor $g_{\mu \nu}$, and $g$ is the metric
determinant. The last term in Eq.~\eqref{eq:full-action} is the action
for all matter degrees of freedom, which couple minimally to the
metric tensor and do not couple to $\vartheta$.

The Chern-Simons correction is mediated by a canonically-normalized
scalar field $\vartheta$, whose kinetic term in the action is
\begin{align}
I_{\vartheta} = \int d^{4}x \sqrt{-g} \left(- \frac{1}{2} \left(\partial_{a} \vartheta\right) \left(\partial^{a} \vartheta\right)\right)\,.
\end{align}
This scalar field couples non-minimally to the metric through the
term in the action
\begin{align}
I_{\cs} = \int d^{4}x \sqrt{-g} \;  \left(- \frac{1}{4} \frac{\alpha}{\kappa} \; \vartheta \; \pont \right)\,,
\label{eq:action-DCS}
\end{align}
where the Pontryagin density is defined via
\begin{align}
\pont := {^\ast\!}R^{abcd} R_{abcd} =
\frac{1}{2} \epsilon^{abef} R_{ef}{}^{cd} \, R_{abcd} \,,
\end{align}
and $\epsilon^{abcd}$ is the Levi-Civita tensor. Notice that the
definition of the Pontryagin density here differs from that
of~\cite{Alexander:2009tp} by a minus sign, which is compensated by an
additional minus sign in $I_{\cs}$.

Variation of the action with respect to the metric yields the field
equations
\begin{align}\label{eq:MetricEOM1}
G_{ab} + 2 \, \alpha \kappa \, C_{ab} &= \kappa^{2} T_{ab}\,,
\end{align}
where $G_{ab}$ is the Einstein tensor, and the traceless `C-tensor' is
defined as
\begin{align}
C^{ab} = \left(\nabla_{c} \vartheta \right) \; \epsilon^{cde(a} \nabla_{e} R^{b)}{}_{d} + \left(\nabla_{c} \nabla_{d} \vartheta \right) \; {}^{\ast}R^{d(ab)c}\,.
\label{eq:C-tensor}
\end{align}
The stress-energy tensor decomposes linearly into a term that depends
only on the matter degrees of freedom and a term that depends only on
the scalar field, i.e.~$T_{ab} = T_{ab}^{\Mat} + T_{ab}^{\vartheta}$,
where the latter is
\begin{align}
T_{ab}^{\vartheta} := \left(\nabla_{a} \vartheta \right) \left( \nabla_{b} \vartheta \right) - \frac{1}{2} g_{ab} \left(\nabla_{c} \vartheta \right) \left(\nabla^{c} \vartheta \right)\,.
\end{align}
Variation of the action with respect to the scalar field yields its
evolution equation
\begin{align}
\square \vartheta &= \frac{\alpha}{4 \kappa} \, \pont\,,
\label{eq:theta-evolution}
\end{align}
where $\square$ stands for the d'Alembertian operator. Notice that
there is no potential associated with the scalar field, which implies
it is a long-ranged field.
This vanishing (or flat) potential means that $\vartheta$ retains a
global shift symmetry, $\vartheta\to\vartheta +$const., because
$\pont$ is related to a topological invariant~\cite{Yagi:2015oca}.
Retaining this shift symmetry may be important to protect against
certain quantum corrections.

The theoretical motivation to study dCS is varied. From a
string-theory standpoint, non-minimal scalar couplings of the form of
Eq.~\eqref{eq:action-DCS} arise in the low-energy limit of heterotic
string theory upon four-dimensional
compactification~\cite{Green:1984sg,Green:1987mn} (for a review
see~\cite{Alexander:2009tp}). From a loop
quantum gravity standpoint, dCS arises when the Barbero-Immirzi
parameter is promoted to a scalar field in the presence of
fermions~\cite{Taveras:2008yf,Mercuri:2009zt}. From a cosmology
standpoint, the interaction in Eq.~\eqref{eq:action-DCS} arises as one
of three terms that remain in an effective field theory treatment of
single-field inflation~\cite{Weinberg:2008hq}.

The choice of conventions made here differs from that
of~\cite{Alexander:2009tp}. The mapping between the two sets is
$\kappa_{\ay} = 1/(2 \kappa^{2})$, $\beta_{\ay} = 1$ and
$\alpha_{\ay} = \alpha/\kappa$. Moreover, we retain all factors of
$G$, or equivalently of $\kappa$, since we do not set $G$ to
unity.  Without requiring the action to have any specific sets of
units, demanding consistency between $I_{\eh}$, $I_{\cs}$ and
$I_{\vartheta}$ implies $[\vartheta] = [\kappa]^{-1}$ and
$[\alpha] = L^{2}$, where $L$ stands for units of length. Given some
GR solution with characteristic length scale $\mathcal{L}$,
corrections are then controlled by the dimensionless parameter
$\zeta := \alpha^{2}/ {\cal{L}}^{4}$. One can see this by noting that
$|\partial_{ab} \vartheta| \propto (\alpha/\kappa) {\cal{L}}^{-4}$
from Eq.~\eqref{eq:theta-evolution}, which implies that
$|C_{ab}| \propto (\alpha/\kappa) {\cal{L}}^{-6}$ from
Eq.~\eqref{eq:C-tensor}. Then the fractional corrections to GR are proportional
to
$(\alpha \kappa |C_{ab}|/|G_{ab}|) \propto \alpha^{2} {\cal{L}}^{-4} =
\zeta$.

Current constraints on dCS are rather weak because dCS corrections
are relevant only in scenarios where the spacetime curvature is
large. One can see this by noting that dCS corrections to the
gravitational field are sourced by the scalar field, which in turn is
only sourced by the spacetime curvature. In fact, one can easily show
through the argument given in the previous paragraph that constraints
on the $\alpha$ parameter of dCS will be roughly proportional to a
power of ${\cal{L}}$. Let us assume that some observation places the
constraints $|\zeta| < \delta$, where $\delta$ is related to the
observation and its uncertainties. This constraint
can then be mapped to a constraint on $\alpha$ to find
$\sqrt{|\alpha|} < \delta^{1/4} {\cal{L}}$. Currently, the best
constraint on the dCS coupling parameter is
$\sqrt{|\alpha|} \lesssim 10^{8} \; {\rm{km}}$ and it comes from
observations of Lense-Thirring precession from satellites in orbit
around Earth~\cite{AliHaimoud:2011fw}. Such a weak constraint makes
sense when one realizes that for these kind of experiments the
characteristic length scale
${\cal{L}} = [R_{\oplus}^{3}/(G M_{\oplus})]^{1/2} \approx 2 \times
10^{8} \; {\rm{km}}$. Binary pulsar observations cannot yet be used
to constrain the theory because modifications to the orbital dynamics
are too weak and couple to the spin of the bodies~\cite{Yagi:2013mbt}. 

The aforementioned theoretical motivations suggest that one treat dCS as an
\emph{effective theory} valid up to some cut-off scale, i.e., the
scale above which higher-order curvature terms in the action cannot be
neglected~\cite{Yunes:2013dva,Berti:2015itd}.  We will here
restrict attention to physical scenarios in which the effective theory
is valid, and since we are interested only in black holes,
this means we restrict attention to those with masses
$G M \gg \sqrt{\alpha}$.  When this is the case, we can work in the
decoupling limit of the theory, i.e.~we perform a perturbative
expansion of the field equations and their solutions in powers of
$\zeta$.  Henceforth, dCS is exclusively treated in the decoupling
limit.

The decoupling limit can be implemented in practice by expanding the
metric tensor and the scalar field in powers of $\zeta$. In this
paper, we will expand the metric and the scalar field as follows:
\begin{align}\label{eq:MetricExpansion0}
g_{ab} &= g_{ab}^{\ZO} + \zeta^{1/2} \; g_{ab}^{\FO} + \zeta \; g_{ab}^{\SO}+ {\cal{O}}(\zeta^{3/2})\,,
\\ \label{eq:ScalarFieldExpansion0}
\vartheta &= \frac{1}{\kappa} \tsf^{\ZO} + \frac{1}{\kappa}\zeta^{1/2} \; \tsf^{\FO} + \frac{1}{\kappa} \zeta \; \tsf^{\SO} + {\cal{O}}(\zeta^{3/2})\,,
\end{align}
where the superscript denotes the order in $\zeta$ of each term.
Notice that a factor of $\kappa^{-1}$ in the expansion for the scalar
field ensures that $\tilde{\vartheta}^{(n)}$ is dimensionless.  As we
are perturbing about $\zeta=0$, our background solution
$(g^{\ZO},\tsf^{\ZO})$ must solve the field equations for GR and a free massless scalar field.  Choosing trivial
initial data for $\tsf^{\ZO}$ gives $\tsf^{\ZO}=0$ at all times, so we
find $(g_{ab}^{\ZO},\tsf^{\ZO})=(g_{ab}^{\gr},0)$ at zeroth order,
where $g_{ab}^{\gr}$ is some known GR solution.  If we next examine
the system at order $\zeta^{1/2}$, we find that $g_{ab}^{\FO}$
satisfies a homogeneous linear equation due to the vanishing of
$\tsf^{\ZO}$.  Therefore, again, trivial initial data gives
$g_{ab}^{\FO}=0$ at all times.

Thus to the order we are working, our expansion is
\begin{align}\label{eq:MetricExpansion}
g_{ab} &= g_{ab}^{\gr} + \zeta \; g_{ab}^{\SO}+ {\cal{O}}(\zeta^{3/2})\,,
\\ \label{eq:ScalarFieldExpansion}
\vartheta &= 0 + \frac{1}{\kappa}\zeta^{1/2} \; \tsf^{\FO} + {\cal{O}}(\zeta^{1})\,.
\end{align}
Henceforth, we will focus on BH solutions, with the
${\cal{O}}(\zeta^{0})$ term in the metric, $g_{ab}^{\gr}$, being
simply the Kerr metric. The ${\cal{O}}(\zeta^{1/2})$ term in the
scalar field, $\vartheta^{\FO}$, 
is sourced by the Kerr metric and, in turn, this sources the
${\cal{O}}(\zeta)$ correction to the metric,
$g_{ab}^{\SO}$.  To be within the regime of validity of the
perturbative expansion, we require $\zeta \ll 1$, and since for the
Kerr black hole the typical curvature length scale is ${\cal{L}} =
GM$, we take
\begin{equation}
\zeta = \frac{\alpha^{2}}{(GM)^{4}} \ll 1\,.
\end{equation}
Notice that this definition differs from others in the
literature~\cite{Alexander:2009tp} in that we do not include a factor
of $1/\kappa^{2}$ in $\zeta$, but rather we factor it out in the
scalar field directly.

In this paper we are concerned with solutions that represent rotating
BHs spinning near extremality, so in addition to the decoupling
expansion we will also perform a \emph{near-extremal
  expansion}. Letting the (z-component of the) BH spin angular momentum be $J_{z}$, we
can define the BH dimensionless spin parameter
$\chi := J_{z}/(G M)^{2}$. We can then expand all fields in the
problem in a bivariate expansion, i.e.~a simultaneous expansion in
both $\zeta \ll 1$ and $\chi \sim 1$, namely
\begin{align}\label{eq:MetricExpansion-extremal}
g_{ab}^{(n)} &= g_{ab}^{(n,0)} + \eps \; g_{ab}^{(n,1)} + \eps^{2} \; g_{ab}^{(n,2)} + {\cal{O}}(\eps^{3})\,,
\\ \label{eq:ScalarFieldExpansion-extremal}
\tsf^{(n)} &=  \tsf^{(n,0)} + \eps \; \tsf^{(n,1)} + \eps^{2} \; \tsf^{(n,2)} + {\cal{O}}(\eps^{3})\,,
\end{align}
where $\eps := (1 - \chi^{2})^{1/2}$ is a near-extremality
parameter and $\eps \ll 1$ for near-extremal BHs.

\section{Scalar Field: Solution}
\label{sec:ScalarFieldSolutions}

We wish to solve the evolution equation for the scalar field
[Eq.~\eqref{eq:theta-evolution}] to leading order in $\zeta$. To this
order, the Pontryagin density on the right-hand side of
Eq.~\eqref{eq:theta-evolution} is evaluated on the unmodified Kerr
spacetime. The wave operator on the left-hand side can also be
evaluated on the Kerr spacetime, since corrections will be of
${\cal{O}}(\zeta)$. In polynomial Boyer-Lindquist coordinates, the scalar field
evolution equation is evaluated on the line element~\cite{Kerr:1963ud}
\begin{align} \label{Kerr}
  g_{ab}^{\ZO} dx^{a} dx^{b} = - \frac{\Delta}{\Sigma} \big[dt - a\, \Gamma \,d\phi\big]^{2} + \frac{\Sigma}{\Delta} dr^{2}  + \frac{\Sigma}{\Gamma}d\notu^2 + \frac{\Gamma}{\Sigma}\,\Big((r^2 + a^2) d\phi - a\,dt\Big)^2 ,
\end{align}
where the usual polar angle $\theta$ has been replaced with a
coordinate $\notu = \cos\theta$, and
$\Gamma := 1 - \notu^{2} = \sin^{2}{\theta}$. The mass of the black
hole is $M$ and it rotates with angular momentum per unit mass
$a = J_{z}/(G M)$, where $-G M \leq a \leq G M$. The functions
$\Sigma$ and $\Delta$ are
\begin{align}
  \Sigma = &\, r^{2} + a^{2} \notu^{2} \\
  \Delta = &\dub r^{2} - 2 G M r + a^{2} ~,
\end{align}
so that the background horizons, where $\Delta = (r-r_{+})(r-r_{-}) = 0$, are
located at $r_{\pm}=GM \pm \sqrt{(GM)^2 - a^2}$.

It will be convenient to replace all quantities with dimensionless
variables by scaling out factors of $G M$: $\TR = r/(G M)$ and 
$\chi = a/(G M)$, so that the rescaled functions
$\tD = \Delta/(G M)^2 = (\TR-1)^2 - (1-\chi^2)$ and
$\tS = \Sigma/(G M)^2 = \TR^2 + \chi^2 \notu^2$. Assuming a stationary
and axisymmetric solution for the scalar field, the
${\cal{O}}(\alpha)$ term in Eq.~\eqref{eq:theta-evolution} then takes
the form
\begin{align} \label{ScalarEOM2}
  \partial_{\TR}\big(\tD \partial_{\TR} \tsf^{\FO} \big) + \partial_{\notu}\big(\Gamma \partial_\notu \tsf^{\FO} \big) = s^{\FO}(\TR,\notu)
\end{align}
where factors of $(\alpha/\kappa)$ and $(G M)$ have canceled from both sides of the equation. The source $s^{\FO}(\TR,\notu)$ is proportional to $\Sigma \; (\pont^{\ZO})$ and given explicitly by
\begin{align} \label{ScalarSource}
  s^{\FO}(\TR,\notu) = 24\,\frac{\chi \TR \notu (3\TR^2 - \chi^2 \notu^2)(\TR^2 - 3\chi^2 \notu^2)}{\tS^{5}} ~.
\end{align}

Equation~\eqref{ScalarEOM2} admits a solution via separation of
variables, by expanding the solution
\begin{align}\label{eq:LegendreDecomp}
  \tsf^{\FO} = \sum_{\ell=0}^{\infty}\tsf_{\ell}^{\FO}(\TR) P_{\ell}(\notu) ~\,,
\end{align}
where $P_{\ell}(\cdot)$ are Legendre functions of the first kind. The
radial modes $\tsf_{\ell}^{\FO}(\TR)$ then satisfy the equation
\begin{align}\label{ScalarEOM3}
  \partial_{\TR}\Big(\tD \partial_{\TR} \tsf^{\FO}_{\ell} \Big) - \ell(\ell+1) \tsf^{\FO}_{\ell} =  s_{\ell}^{\FO}(\TR)~,
\end{align}
with source functions $s_{\ell}^{\FO}(\TR)$ given by the modes in the
Legendre decomposition of Eq.~\eqref{ScalarSource}
\begin{align}\label{ScalarSourceFunctions}
  s_{\ell}^{\FO}(\TR) = \frac{2\ell+1}{2} \! \int_{-1}^{1}\nts \nts d\notu\, P_{\ell}(\notu) s^{\FO}(\TR,\notu) ~.
\end{align}
Note that the source function in Eq.~\eqref{ScalarSource} is odd in
the variable $\notu$, so its Legendre expansion (as well as that of
the scalar field) will only contain odd modes: $\ell=2n+1$ for all
$n \in {\mathbb{N}}$.

The integral in Eq.~\eqref{ScalarSourceFunctions} can be evaluated in
closed form in terms of known functions:
\begin{multline}
s_{\ell}^{\FO}(\TR) = (-1)^{\frac{\ell+1}{2}}\,\frac{\Gamma(\tfrac{1}{2})\Gamma(\ell+4)}{2^{\ell} \Gamma(\ell+\tfrac{1}{2})}\,\frac{\chi^{\ell}}{\TR^{\ell+4}}\,
 \left[\; 3\, {}_{2}F_{1}\left(\textstyle\frac{\ell+4}{2},\frac{\ell+5}{2};\ell+\frac{3}{2};-\frac{\chi^{2}}{\TR^{2}}\right)
\right.
\\ \left.
-(\ell+5) \; {}_{2}F_{1}\left(\textstyle\frac{\ell+4}{2},\frac{\ell+7}{2};\ell+\frac{3}{2};-\frac{\chi^{2}}{\TR^{2}}\right) \right]\,,
\label{eq:nasty-long-source}
\end{multline}
where $_{2}F_{1}(\cdot,\cdot;\cdot;\cdot)$ is the ordinary
hypergeometric function and $\ell$ is odd. One can show, via
identities for hypergeometric functions, that this expression is
equivalent to one given previously in~\cite{Stein:2014wza}. Note that the hypergeometric functions go to unity in the $\TR \to \infty$ limit, so that the leading behavior at large $\TR$ is given by $s_{\ell}^{\FO}(\TR) \sim \TR^{-(\ell+4)}$. 

The solution of Eq.~\eqref{ScalarEOM3} can be obtained through the
method of variation of parameters (see~\ref{app:GeneralSolution}). Defining a new variable
$\eta = (\TR-1)/\sqrt{1-\chi^2}$, the solution of
Eq.~\eqref{ScalarEOM3} [see also
Eq.~\eqref{eq:GeneralScalarEOMSolution}] for the mode function
$\tsf^{\FO}_{\ell}$ is
\begin{multline} \label{NESolution}
  \tsf_{\ell}^{\FO}(\TR) = \, P_{\ell}(\eta) \int_{\infty}^{\eta} d\eta' \,s_{\ell}^{\FO}(1 + \eta' \sqrt{1-\chi^2}) \; Q_{\ell}(\eta') \\
    {}- Q_{\ell}(\eta) \int_{1}^{\eta} \! d\eta' \,s_{\ell}^{\FO}(1 + \eta'\sqrt{1-\chi^2}) \; P_{\ell}(\eta') \,,
\end{multline}
where $Q_{\ell}(\cdot)$ are Legendre functions of the second
kind. This solution is regular at $\TR_{+}$, and approaches zero as
$\TR \to \infty$.

Our eventual goal is to evaluate Eq.~\eqref{NESolution} in closed form
for the full range of the rotation parameter, $-1 \leq \chi \leq 1$.
The slow rotation limit of the field, i.e.~the solution in a
$|\chi| \ll 1$ expansion, is already well-understood; it was first
derived in~\cite{Yunes:2009hc}, verified in~\cite{Konno:2009kg}, and
extended to second order in rotation in~\cite{Yagi:2012ya}. Similarly,
it is also possible to systematically solve the scalar field equation
of motion in the near-extremal expansion introduced in
Sec.~\ref{sec:ABC-DCS}. Expanding the source functions of
Eq.~\eqref{eq:nasty-long-source} for $\eps \ll 1$, we find
\begin{align}
  s_{\ell}^{\FO}(\TR) = s_{\ell}^{\FOZO}(\TR) + \eps^{2} s_{\ell}^{\FOSO}(\TR) + \eps^{4} s_{\ell}^{\FOFFO}(\TR) + {\cal{O}}(\eps^{6})\,.
\end{align}
Recall that the second superscript in each of these terms represents
the order in $\eps$ at which it enters the near-extremal expansion.
Because the $\chi\to 1$ limit is regular for $s_{\ell}^{\FO}(\TR)$,
$s_{\ell}^{\FOZO}(\TR)$ is simply $s_{\ell}^{\FO}(\TR)$ evaluated at
$\chi=1$.

In this paper we will only consider the extremal limit,
$\eps \to 0$, which is the leading term in the near-extremal
expansion. This corresponds to the limit $\chi \to \pm 1$ of the
dimensionless spin parameter $\chi$. The homogeneous solutions are
regular in this limit [see
Eqs.~\eqref{eq:HplusExt}-\eqref{eq:HminusExt}], and the solution for
the scalar field [see Eq.~\eqref{eq:GeneralScalarEOMSolution}] at
$\eps =0$ is
\begin{multline}\label{eq:extremal-theta}
  \tsf^{\FOZO}_{\ell}(\TR) = \, \frac{1}{2\ell+1}\left[(\TR-1)^{\ell} \int_{\infty}^{\TR}
  \! d\TR'\,\frac{s_{\ell}^{\FOZO}(\TR')}{(\TR'-1)^{\ell+1}}\right. \\
  - \left. \frac{1}{(\TR-1)^{\ell+1}} \int_{1}^{\TR} \!
  d\TR'\,(\TR'-1)^{\ell}s_{\ell}^{\FOZO}(\TR')\right] ~.
\end{multline}
The source functions at leading-order in $\eps$,
$s^{\FOZO}_{\ell}(\TR)$, are given by
Eq.~\eqref{ScalarSourceFunctions} or Eq.~\eqref{eq:nasty-long-source}
evaluated at $|\chi| = 1$. The boundary conditions are the same as
before: each mode $\tsf^{\FOZO}_{\ell}$ is regular at $\TR_{+} = 1$,
and goes to zero as $\TR \to \infty$.

The integrals in Eq.~\eqref{eq:extremal-theta} can be readily
evaluated for specific values of $\ell$. For example, the $\ell=1$
radial mode is given by
\begin{align}\nn
  \tsf^{\FOZO}_{1}(\TR) = &\dub 3(\TR-1)\log\Big(\frac{\TR-1}{\sqrt{\TR^2+1}}\Big) + 3 (\TR-1)\arccot\TR \\  \label{eq:FirstModeYay}
  & {}+ \frac{3(\TR-1)(2\TR^2 + \TR + 3)}{2(\TR^2+1)^2} .
\end{align}
With more work, we can also give an expression for general values of
$\ell$ in terms of finite-order rational polynomials, $\arccot(\TR)$,
and the $\log$ which appears above.  But before we can give the general
form, we first have to establish a few results for the behavior of the
modes at large $\TR$ and at $\TR = 1$. 

The far-field behavior of the modes is dominated by the second line of
Eq.~\eqref{eq:extremal-theta}, since the term in the first line decays with a
higher power of $r$.
The integral in the second line converges in this
limit, and thus, $\tsf^{\FOZO}_{\ell} \sim \TR^{-(\ell+1)}$ as
$\TR \gg 1$. Notice, though, that this asymptotic behavior is not immediately apparent in our initial result, Eq.~\eqref{eq:FirstModeYay}. In that case the individual terms fall off more slowly than $\TR^{-2}$, but cancellations between the terms result in the correct asymptotic behavior. The same will be true for our result for general $\ell$: individual terms may not behave as $\TR^{-(\ell+1)}$ at large $\TR$, but cancellations between these terms will give the correct result.

The near-horizon behavior of the modes is dominated by
the first integral in Eq.~\eqref{eq:extremal-theta}, but its
asymptotic behavior as $\TR \sim 1$ cannot be easily discerned from
that equation. Instead, it is easier to return to
Eq.~\eqref{NESolution} and set $\chi =1$, remembering that the horizon
limit $\TR \to 1$ is equivalent to $\eta \to 1$ (this is the case for
all values of $\chi$). In this limit, only the first line of
Eq.~\eqref{NESolution} contributes, leading to
\begin{align}\label{eq:horizon-value}
  \tsf_{\ell}^{\FOZO}(1) = & \dub -\frac{1}{\ell(\ell+1)}\,s_{\ell}^{\FOZO}(1) ~.
\end{align}
The overall $\ell$-dependent factor comes from the definite
integration of $Q_{\ell}(\eta')$ in the range $\eta' \in (1,\infty)$,
while $s_{\ell}^{\FOZO}(1)$ is Eq.~\eqref{eq:nasty-long-source}
evaluated at $\chi=1$ and $\TR=1$. As a check of this result, it is
instructive to consider the behavior of the scalar field in the
Near-Horizon Extremal Kerr (NHEK)
limit~\cite{Bardeen:1999px,Guica:2008mu}. At extremality, regularity
of the field at the horizon insures that the first term on the
left-hand-side of Eq.~\eqref{ScalarEOM2} vanishes as $\TR \to 1$. Then
the $\notu$-dependence of the scalar field in this limit is determined
by
\begin{gather}\label{HorizonEOM}
  \partial_{\notu}\big(\Gamma \partial_\notu \tsf^{\FOZO}(1,\notu) \big) = s^{\FOZO}(1,\notu) = \frac{24\,\TR\,\notu\,(3-\notu^2)(1-3\notu^2)}{(1 + \notu^2)^5} ~.
\end{gather}
The inhomogeneous solution can be obtained by direct integration, and the full solution that is regular on $-1\leq \notu \leq 1$ is given by
\begin{gather}\label{HorizonField}
	\tsf^{\FOZO}(1,\notu) = -\frac{4\,\notu}{(1+\notu^2)^3} + \frac{\notu}{2(1+\notu^2)} + \arctan(\notu) ~.
\end{gather}
It is straightforward to check that this agrees with the Legendre series Eq.~\eqref{eq:LegendreDecomp}, with coefficients given by Eq.~\eqref{eq:horizon-value} for the mode functions at the horizon.

With these results, we can now give a general expression for the
radial modes of the scalar field. They take the form
\begin{align}\label{eq:General-Mode}
  \tsf^{\FOZO}_{\ell}(\TR) = & \,\, A_{\ell}(\TR) + B_{\ell}(\TR)\,\arccot(\TR)+ C_{\ell}(\TR)\, \log\left(\frac{\TR-1}{\sqrt{\TR^2 + 1}}\right) ~,
\end{align}
where the functions $A_{\ell}(\TR)$, $B_{\ell}(\TR)$, and
$C_{\ell}(\TR)$ are
\begin{align}
  A_{\ell}(\TR) = & \,\, (-1)^{\frac{\ell-1}{2}}\,\frac{\ell(\ell-1)}{(\TR-1)^{\ell+1}}\,\sum_{k=0}^{\ell} \gamma_{k}\,(\TR-1)^{k} + \sum_{k=0}^{\ell-1} \alpha_{\ell,k}\,\TR^{k} \\ \nonumber
  & \quad - \frac{2\ell+1}{(\TR^2 + 1)^2} + \frac{(2\ell+1)(4\TR - \ell(\ell+1))}{4\,(\TR^2 + 1)} \\
  B_{\ell}(\TR) = & \,\, (-1)^{\frac{\ell+1}{2}}\,\frac{\ell(\ell-1)}{(\TR-1)^{\ell+1}} + \sum_{k=0}^{\ell} \beta_{\ell,k}\,\TR^{k} \\
  C_{\ell}(\TR) = & (-1)^{\frac{\ell-1}{2}} \frac{(\ell+1)(\ell+2)}{2}\,(\TR-1)^{\ell}.
\end{align}
The constants $\gamma_{k}$ appearing in $A_{\ell}(\TR)$ are the first
$\ell+1$ terms in the Taylor expansion of $\arccot(\TR)$ around
$\TR=1$
\begin{align}
  \gamma_{k} = \frac{1}{k!}\,\left(\frac{\partial}{\partial x^{k}} \arccot (x)\right)\Bigg|_{x=1}~.
\end{align}
The remaining $2\ell+1$ coefficients $\alpha_{\ell,k}$ and
$\beta_{\ell,k}$ are fixed by imposing the boundary conditions: each
mode falls off as $\TR^{-(\ell+1)}$ at large $\TR$, and takes the
value Eq.~\eqref{eq:horizon-value} at $\TR=1$. Alternately, the
condition at $\TR=1$ can be replaced with the requirement that the
leading asymptotic behavior of the mode is given by
Eq.~\eqref{eq:aL0}. The coefficients $\alpha_{\ell,k}$ and
$\beta_{\ell,k}$ for the first several modes are given explicitly in~\ref{app:higher-modes}.

\section{Scalar Field: Properties}
\label{sec:ScalarFieldProperties}

\begin{figure}[t]
  \centering
  \includegraphics[width=0.9\columnwidth{}]{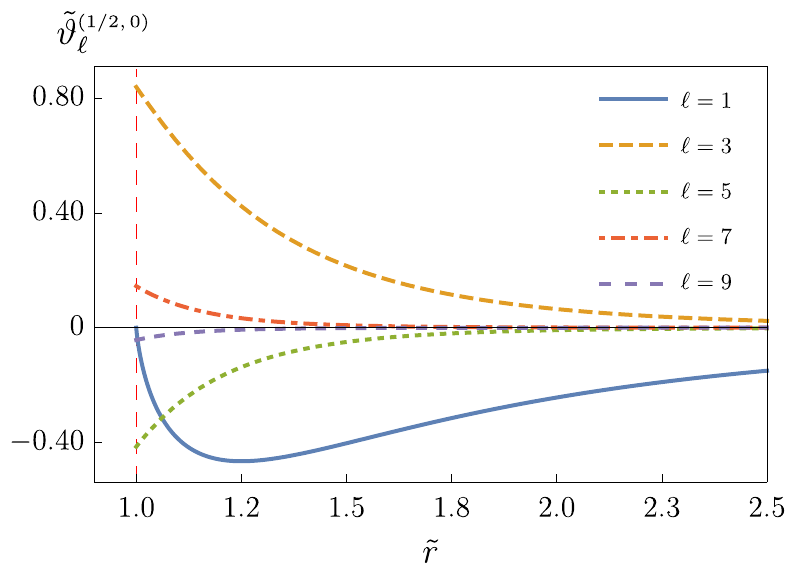}
  \caption{%
    The first five radial mode functions of the scalar field. The
    vertical dashed line indicates the location of the event horizon
    of an extremal black hole. The $\ell=1$ mode vanishes at the
    horizon, while modes with $\ell > 1$ are all non-zero at
    $\TR = 1$.  }
  \label{fig:ModeFunctions}
\end{figure}

Let us now discuss some properties of the scalar field solution
obtained in the previous section. We begin by plotting the first five
(odd) modes in Fig.~\ref{fig:ModeFunctions}. Observe that the
integrated norm of $\tsf^{\FOZO}_{\ell}$ decays exponentially with
$\ell$. This is because this function is a spectral solution to a
differential equation with a $C^{\infty}$ source, so it must converge
exponentially with mode number.  Observe also from
Fig.~\ref{fig:ModeFunctions} that the $\ell=1$ mode of the field
vanishes at the horizon. Modes with $\ell > 1$ are finite but non-zero
at the horizon, with values that scale like
$\ell^{5/2}(1+\sqrt{2})^{-(\ell+1)}$ for $\ell \gg 1$.

\begin{figure}[t]
  \centering
  \includegraphics[width=0.8\columnwidth{}]{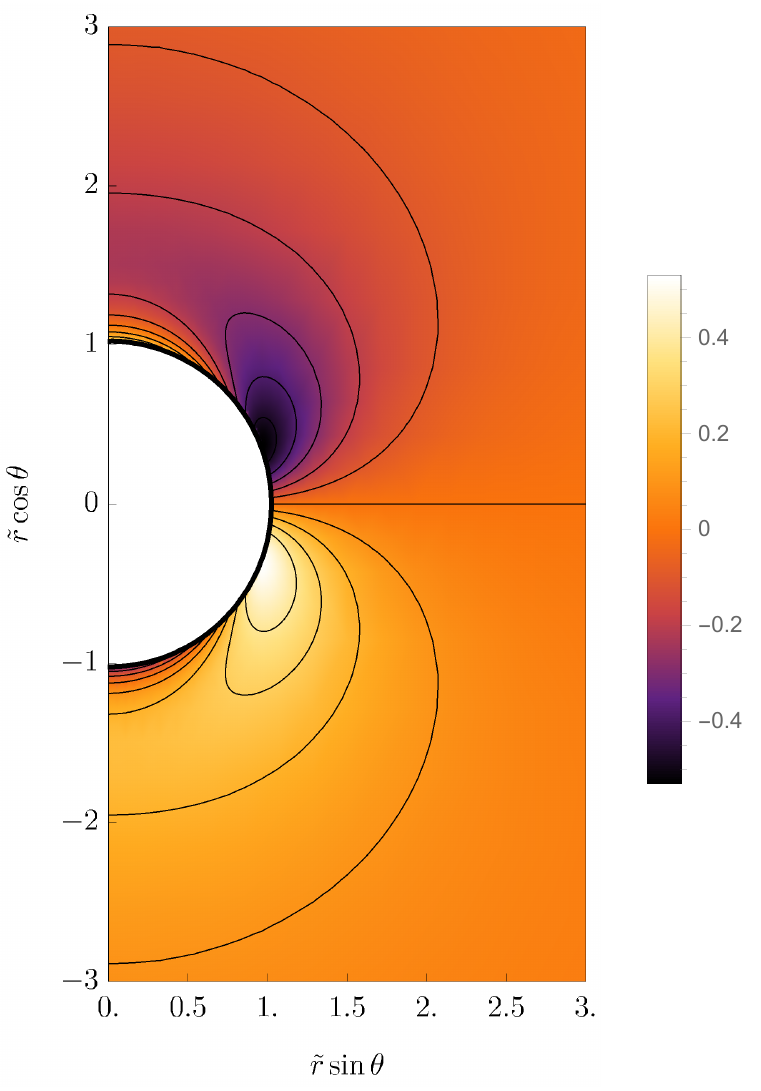}
  \caption{%
The behavior of the scalar field on the extremal background,
approximated by its first five Legendre modes. The coordinates
$\TR\notu$ and $\TR\sqrt{1-\notu^{2}}$ correspond to $\TR\cos\theta$
and $\TR \sin\theta$, respectively, in conventional Boyer-Lindquist
coordinates.
  }
  \label{fig:ScalarDensityPlot}
\end{figure}

By including contributions from a sufficient number of modes, we can
construct an arbitrarily accurate approximation of the full, extremal
scalar field. An approximation using the first five modes is shown in
Fig.~\ref{fig:ScalarDensityPlot}, as a function of both radius and
polar angle.
Notice the similarity to Fig.~3 of Stein~\cite{Stein:2014xba},
resulting from a numerical solution at large but not extremal
($\chi=0.999$) rotation; we will discuss this more below.
The accuracy of this approximation can be characterized using a
slicing-independent measure of the scalar field energy through the ADM
energy. Let $u^{a}$ be a timelike unit vector normal to a hypersurface
$\mathcal{S}$, with $\gamma_{ab}$ the induced metric on
$\mathcal{S}$. Then the scalar field's contribution to the energy is
\begin{gather}
  E = \int_{\mathcal{S}} d^{3}x \; \sqrt{\gamma} \; u^{a} T_{ab}^{\vartheta} t^{b}
\end{gather}
where $t^{b}$ is the Killing vector $\partial/\partial t$. This energy
can be perturbatively expanded in powers of $\zeta$,
\begin{gather}
  E = \zeta\,E^{(1)} + \zeta^{2}\,E^{(2)} + \ldots
\end{gather}
The scalar field's ADM energy at leading order can further be computed
via the spectral decomposition,
\begin{gather}
  E^{\SO} = M \sum_{k=1}^{\infty} \tilde{E}_{k}^{\SO},
\end{gather}
with the dimensionless $\tilde{E}_{k}^{\SO}$ functions given by
\begin{align}
\label{eq:tildeEkSO}
\tilde{E}_{k}^{\SO} &= \frac{1}{4} \frac{1}{2k+1} \int_{\TR_{+}}^{\infty} d\TR \Big[\tD (\partial_{\TR} \tsf_{k}^{\FO})^{2} + k(k+1) (\tsf_{k}^{\FO})^{2}\Big] ~.
\end{align}
The fractional difference between the total energy in the scalar field and the energy in the first $N$ modes is then
\begin{gather}
  \delta_{N} = 1 - \frac{M}{E^{\SO}} \sum_{k=1}^{N} \tilde{E}_{k}^{\SO} ~.
\end{gather}

\begin{figure}[t]
  \centering
  \includegraphics[width=0.9\columnwidth{}]{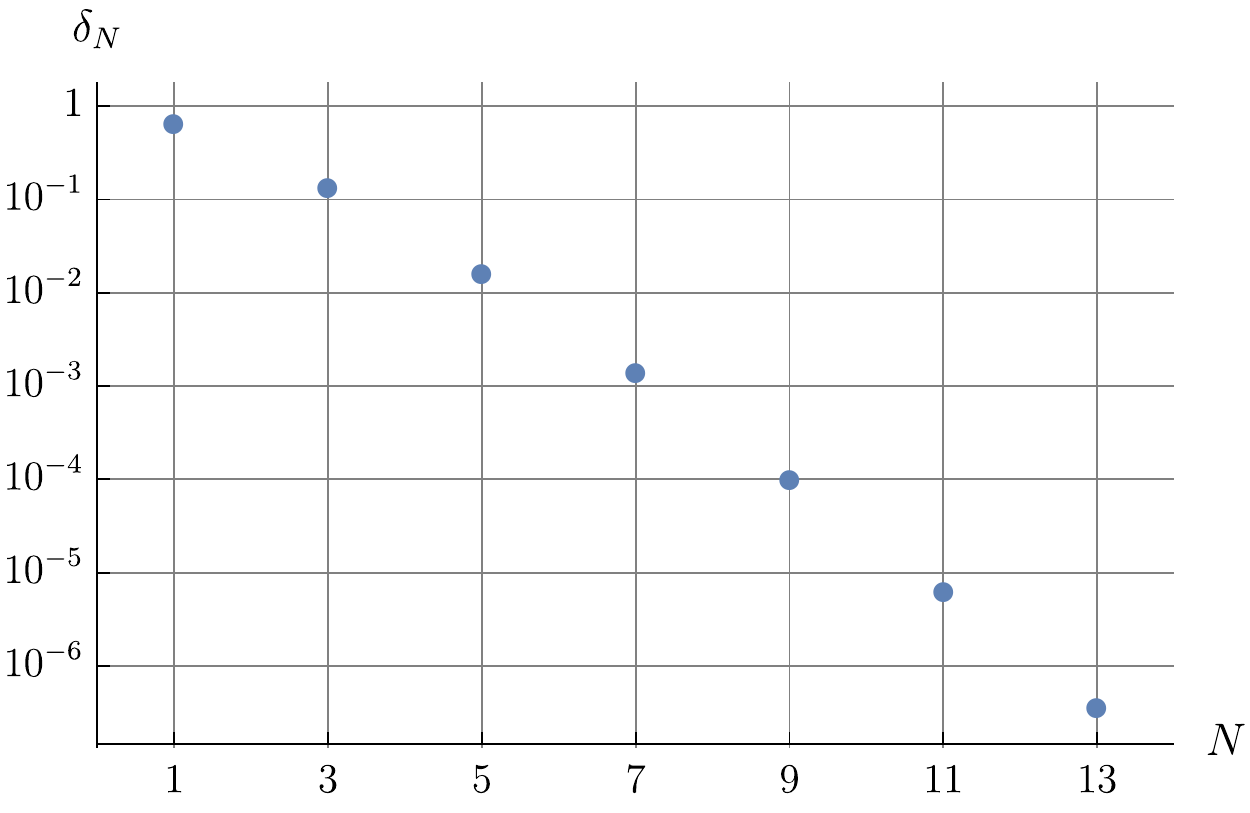}
  \caption{%
The fractional difference between the scalar field's contribution to the ADM energy, and the contribution from the field's first $N$ modes.
  }
  \label{fig:deltaN}
\end{figure}

Figure~\ref{fig:deltaN} shows the fractional difference $\delta_{N}$
for the first seven nonvanishing modes.  The contribution from the first five
-- up to $\ell=9$ -- differs from the total energy by less than one
part in $10^{4}$. Observe that the accuracy increases exponentially
with $N$. Observe also that if we wish to capture $99\%$ of the energy
in the field, it suffices to keep only up to the first three odd modes,
i.e.~$N=5$. Finally, note that the energy in the scalar field is
dominated by the behavior of the scalar field close to the horizon. If
one is interested in regimes of spacetime outside some two-sphere with
radius $r \gg M$, then the full scalar field can be accurately
modeled using just the dipole ($\ell=1$) and octupole ($\ell=3$)
modes.

\begin{figure}[tbp]
  \centering
  \includegraphics[width=0.9\columnwidth{}]{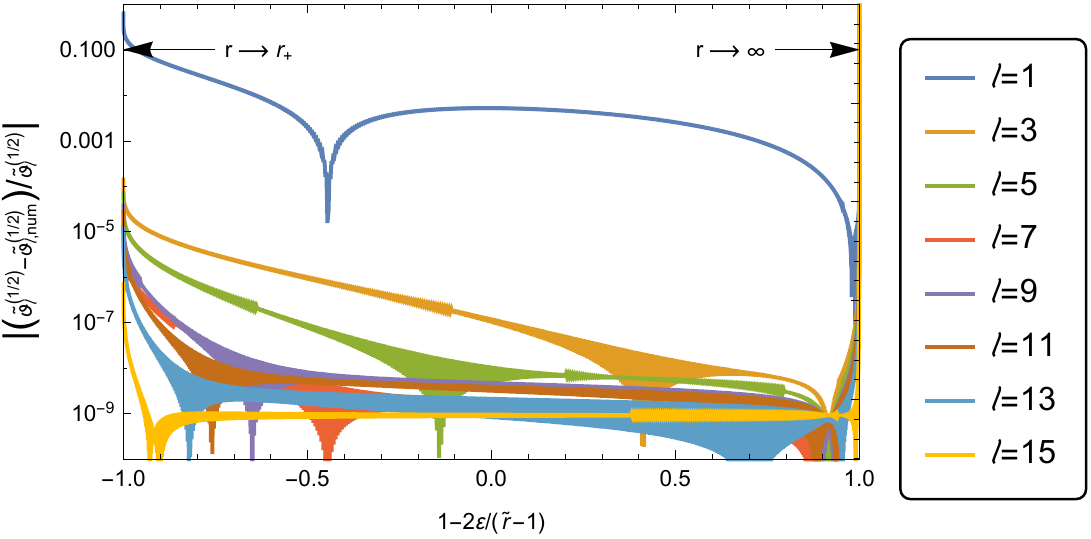}
  \caption{%
Direct comparison between the analytical solution
$\tilde{\vartheta}_{\ell}^{\FO}$ and the numerical solution at
$a=1-10^{-10}$ (approximately $\varepsilon \approx \sqrt{2}\times
10^{-5}$), using the method of~\cite{Stein:2014xba}. The vertical axis
is fractional difference between the numerical solution and the
analytical solution for each mode. The extremal
solution is evaluated onto the collocation points of the numerical
domain $x\in(-1,+1)$ where $x=1-2\varepsilon/(\tilde{r}-1)$. The
numerical solution was computed with an angular and radial resolution
of $(N_{ang},N_{x})=(64,1024)$, and comparisons against lower
resolutions shows that the numerics are converging. The limit
$\varepsilon\to 0$ is regular, so the fractional differences will go
to zero.
  }
  \label{fig:extremalFracDiffPlot}
\end{figure}

Let us briefly compare these analytic results at extremality with
numerical results away from extremality, which were computed in
Stein~\cite{Stein:2014xba}. First, we have performed a direct
comparison of the analytical calcuation presented here with the
numerics in~\cite{Stein:2014xba}. Because of the numerical method
of~\cite{Stein:2014xba}, those results must be at values of
$|\chi|<1$. Specifically, we compared the analytical results presented
here with numerical results computed with the highest spin of
$\chi = 1 - 10^{-10}$ (extremality parameter approximately
$\varepsilon \approx \sqrt{2}\times 10^{-5}$), with $N_{ang}=64$
angular and $N_{x}=1024$ radial collocation points (high spins require
very high radial resolution). Comparisons against lower resolutions
show that these results are converging.  This comparison is displayed
in Fig.~\ref{fig:extremalFracDiffPlot}. We see that even at the
extremality parameter $\varepsilon \approx \sqrt{2}\times 10^{-5}$,
the numerical solution is converging to the regular analytic
solution. The nonzero fractional differences in
Fig.~\ref{fig:extremalFracDiffPlot} should decrease in the limit as
$\varepsilon \to 0$ and in the continuum limit $N_{x}\to\infty$.

\begin{figure}[tbp]
  \centering
  \includegraphics[height=5.7cm]{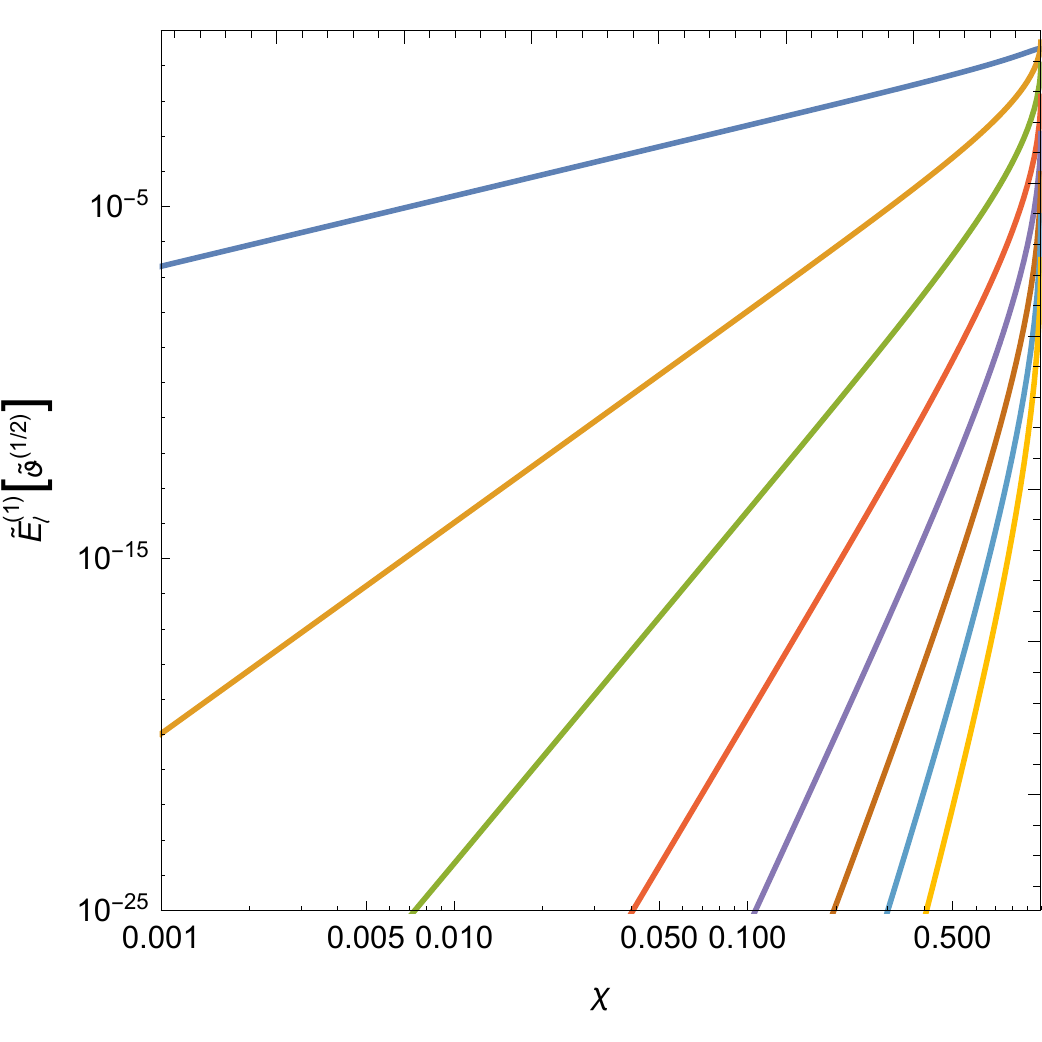}
  \includegraphics[height=5.7cm]{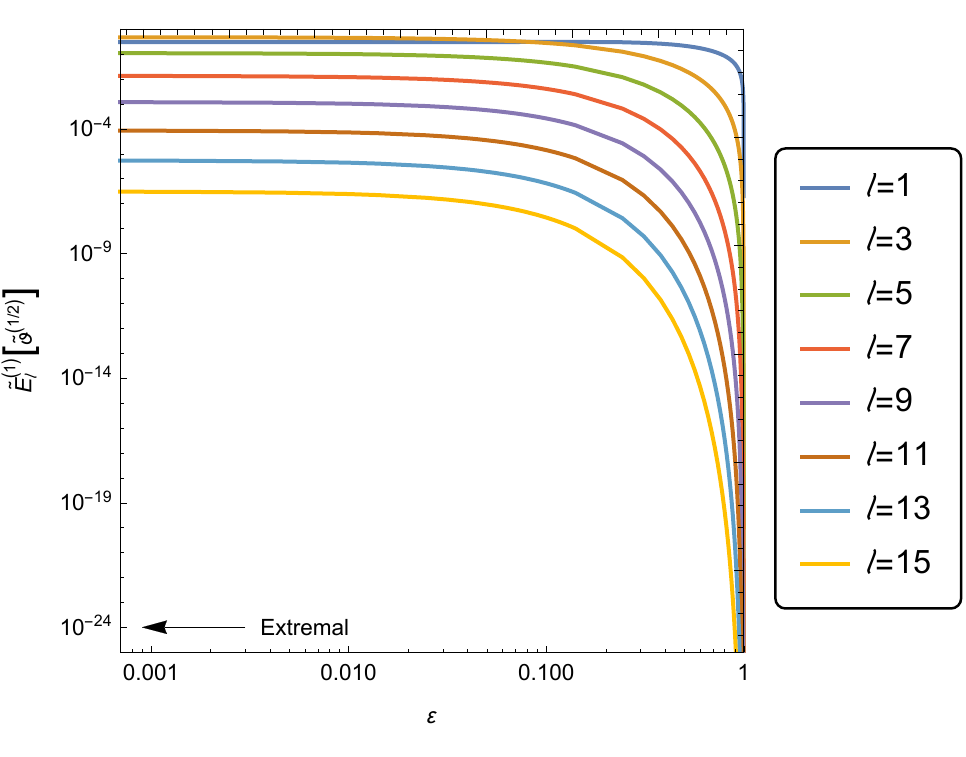}
  \caption{%
Modal contribution to the ADM energy of the scalar field [defined in
Eq.~\eqref{eq:tildeEkSO}], as a function of rotation on the horizontal
axis.  Notice that the extremal limit is regular. This is emphasized
by making the horizontal axis the extremality parameter $\varepsilon
= \sqrt{1-\chi^{2}}$. Note that each curve becomes horizontal
going toward the left edge of the right panel.
  }
  \label{fig:ElsthetaCSofa}
\end{figure}

We can also use the numerical method of~\cite{Stein:2014xba} to show
convergence to the extremal limit, which is regular. This is
demonstrated in Fig.~\ref{fig:ElsthetaCSofa} as a function of the
rotation of the background spacetime, by using the modal contribution
to the ADM energy as a proxy. The convergence to the extremal
limit is more easily demonstrated by using the extremality parameter
as the horizontal axis, as seen in the right panel. We can also easily
see that higher $\ell$ modes become fractionally more important as
spin increases (this was seen in Fig.~2 of~\cite{Stein:2014xba}). 

The essential differences between the structure of the scalar field in
the slowly rotating ($\chi \to 0$) and extremal ($\chi \to 1$) limits
are apparent in Fig.~\ref{fig:ElsthetaCSofa}. At low spin, $\chi \ll
1$, the scalar field is almost entirely dipolar, with contributions to
the ADM energy from the $\ell \geq 3$ modes amounting to less than 1
part in $10^{12}$. The higher modes become more important as the spin
of the background black hole increases, and the octupole contribution
to the ADM energy exceeds the dipole contribution at around $\chi
\approx 0.995$. At $\chi = 1$, the dipole contributes about $30\%$ of
the total ADM energy, while the octupole contributes just over
$50\%$. This is consistent with the near-horizon behavior of the modes
shown in Fig.~\ref{fig:ModeFunctions}, and our earlier observation
that the scalar field ADM energy is dominated by the contribution from
the near-horizon region.

\section{Trace of the Metric Perturbation: Solution}
\label{sec:TraceMetPertSolution}

The leading correction to the metric in Eq.~\eqref{eq:MetricExpansion}
is determined by the ${\cal{O}}(\zeta)$ term in the metric equation of
motion [Eq.~\eqref{eq:MetricEOM1}]. We work in a gauge where the
covariant divergence of $g_{ab}^{\SO}$ is proportional to the
derivative of its trace $g^{\SO} = g_{\ZO}^{ab} g_{ab}^{\SO}$ with
respect to the background metric:
\begin{align}\label{eq:LorenzGauge}
  \cd^{a} g_{ab}^{\SO} = \frac{1}{2}\,\nabla_{b}g^{\SO} ~.
\end{align}
This gauge leads to simplifications in the $\mathcal{O}(\zeta)$ term
of Eq.~\eqref{eq:MetricEOM1}, but its structure is still too
complicated to allow for a simple solution. As a first step towards
determining $g^{\SO}_{ab}$, we take
the trace of the $\mathcal{O}(\zeta)$ correction to
Eq.~\eqref{eq:MetricEOM1} to find
\begin{align}\label{eq:Pre-eom-tildeg2}
  \square g^{\SO} = - 2 (\nabla\vartheta^{\FO})^2~.
\end{align}
Assuming a stationary and axisymmetric solution and transforming to
dimensionless variables, this reduces to
\begin{align}
  \label{eq:eom-tildeg2}
  \left[
    \partial_{\TR} \tD \partial_{\TR} + \partial_{\notu} \Gamma \partial_{\notu}
  \right]\tg^{\SO} =
-2\tD (\partial_{\TR}\tsf^{\FO})^{2} - 2 \Gamma (\partial_{\notu}\tsf^{\FO})^2 \,.
\end{align}

As with the scalar field, we can express $\tg^{\SO}$ in a Legendre
decomposition as
\begin{equation}
\tg^{\SO} = \sum_{\ell} \tg^{\SO}_{\ell}(\TR) \; P_{\ell}(\notu)\,.
\end{equation}
Then, the equation of motion [Eq.~\eqref{eq:eom-tildeg2}] again
separates, giving the radial equation
\begin{gather}\label{eq:glModeEquation}
    \left[
      \partial_{\TR} \tD \partial_{\TR} - \ell(\ell+1)
    \right]\tg_{\ell}^{\SO}(\TR) = S_{\ell}(\TR)~.
\end{gather}
The source functions $S_{\ell}(\TR)$ are the Legendre modes of the
right-hand side of Eq.~\eqref{eq:Pre-eom-tildeg2}, i.e.
\begin{gather} \label{eq:SLdefinition}
  S_{\ell}(\TR) = \frac{2\ell+1}{2}\int_{-1}^{1}d\notu P_{\ell}(\notu) S_{g}(\TR,\notu)\,,
\end{gather}
where the source function $S_{g}$ is simply
\begin{equation}
S_{g}(\TR,\notu) := - 2 \sqrt{-\tg^{\ZO}} \; (\nabla\tsf^{\FO})^2\,.
\end{equation}
The solution for the mode functions $g_{\ell}^{\SO}(\TR)$ is then
given by Eq.~\eqref{eq:GeneralScalarEOMSolution}, which in this case
becomes
\begin{equation}\label{eq:g2-FirstSolution}
  \tg_{\ell}^{\SO} = \frac{1}{W_{\ell}}\,\left(H_{\ell}^{+}(\TR)\,\int_{\infty}^{\TR} d\TR' H_{\ell}^{-}(\TR') S_{\ell}(\TR') - H_{\ell}^{-}(\TR) \int_{\TR_{+}}^{\TR} d\TR' H_{\ell}^{+}(\TR') S_{\ell}(\TR') \right)
\,.
\end{equation}
Note that the source is quadratic in the scalar field, which has odd
Legendre modes. Thus, both the trace of the metric perturbation and
its source function have even Legendre modes: $\ell=2n$ for all
$n \in {\mathbb{N}}$.

One approach to evaluating the integrals in
Eq.~\eqref{eq:g2-FirstSolution} is to express the Legendre modes of
the source function in terms of the scalar field modes and their
radial derivatives. The resulting integrals are significantly more
complicated than the ones we encountered in
Sec.~\ref{sec:ScalarFieldSolutions}, so we will opt for a different
approach. One can express Eq.~\eqref{eq:g2-FirstSolution} in terms of
a simpler set of integrals through multiple integrations-by-parts
(noting that the source $S_{\ell}$ depends on $S_{g}$, which in turn
is proportional to the squared derivative of the scalar field) and
application of the scalar field evolution equation
[Eq.~\eqref{ScalarEOM2}]. Doing so, the modes of the trace of the
metric perturbation are given by
\bw
\begin{align}\label{eq:g2Solution2}
\tg^{\SO}_{\ell}(\TR) = \frac{2\ell+1}{W_{\ell}}&\left(
H_{\ell}^{+}(\TR) \int_{\infty}^{\TR} \!\! d\TR' \!\! \int_{-1}^{1} \!\! d\notu \; H_{\ell}^{-}(\TR') P_{\ell}(\notu) \tsf^{\FO}(\TR',\notu) s(\TR',\notu)
\nn \right. \\
& \quad -
\left.  H_{\ell}^{-}(\TR) \int_{\TR_{+}}^{\TR} \!\! d\TR' \!\! \int_{-1}^{1} \!\! d\notu \; H_{\ell}^{+}(\TR') P_{\ell}(\notu) \tsf^{\FO}(\TR',\notu) s(\TR',\notu)
\nn \right. \\
& \quad +
\left. H_{\ell}^{+}(\TR) \int_{\infty}^{\TR} \!\!d\TR' \!\! \int_{-1}^{1} \!\!d\notu \; \partial_{\mu} V^{\mu}_{-}(\TR',\notu)
- H_{\ell}^{-}(\TR) \int_{\TR_{+}}^{\TR}\!\! d\TR' \!\! \int_{-1}^{1} \!\!d\notu \; \partial_{\mu} V^{\mu}_{+}(\TR',\notu) \right)\,,
\end{align}
\ew
where $s(\TR',\notu)$ is the scalar field source given in
Eq.~\eqref{ScalarSource}, and we have defined
\begin{align}
V^{\mu}_{\pm} :=
\frac{1}{2} (\tsf^{\FO})^{2} \; \sqrt{-\tg^{\ZO}} \; \tg_{\ZO}^{\mu \nu} \partial_{\nu} \left[H_{\ell}^{\pm} P_{\ell} \right] -
\frac{1}{2} H_{\ell}^{\pm} P_{\ell} \; \sqrt{-\tg^{\ZO}} \; \tg_{\ZO}^{\mu \nu} \partial_{\nu} (\tsf^{\FO})^{2}\,.
\end{align}

The integrals of total derivatives in Eq.~\eqref{eq:g2Solution2} can
be simplified by noting that (i)
$\sqrt{-\tg^{\ZO}} \tg_{\ZO}^{\notu \notu} = {\Gamma}$, which vanishes
when evaluated at the limits of integration $\notu = \pm 1$, and (ii)
contributions at spatial infinity and at the horizon vanish due to the
behavior of the scalar field modes $\tsf^{\FO}$, the homogeneous
solutions $H_{\ell}^{\pm}$, and
$\sqrt{-\tg^{\ZO}} \tg_{\ZO}^{rr} = \tD$. The modes of the trace of
the metric perturbation are then
\begin{align}\label{eq:gLSolution}
\tg^{\SO}_{\ell}(\TR) &=
\frac{2\ell+1}{W_{\ell}} \Bigg(H_{\ell}^{+}(\TR) \int_{\infty}^{\TR} \!\! d\TR' \!\! \int_{-1}^{1} \!\! d\notu \; H_{\ell}^{-}(\TR') P_{\ell}(\notu) \tsf^{\FO}{} s{}^{\FO}
\nn \\
&-
H_{\ell}^{-}(\TR) \int_{\TR_{+}}^{\TR} \!\! d\TR' \!\! \int_{-1}^{1} \!\! d\notu \; H_{\ell}^{+}(\TR') P_{\ell}(\notu) \tsf^{\FO}{} s^{\FO} \Bigg)
\nn \\
&-
\frac{2\ell+1}{2} \int_{-1}^{1} d\notu \; P_{\ell}(\notu) \; \left(\tsf^{\FO}{}(\TR,\notu)\right){}^{2}\,,
\end{align}
where in the first and second lines $\tsf^{\FO}{}$ and $s^{\FO}$ are
both functions of $\TR'$ and $\notu$. We have simplified the last line
by extracting a factor of $W_{\ell}$, defined in
Eq.~\eqref{Wronskian}, which is a constant.

Let us now focus on the extremal limit. With the normalizations
defined in~\ref{app:GeneralSolution}, the factor
$W_{\ell} = 2 \ell + 1$ and the homogeneous solutions are given by
Eqs.~\eqref{eq:HplusExt}-\eqref{eq:HminusExt}. We can then write
\begin{align}\label{eq:g2ExtSolution}
\tg^{\SOZO}_{\ell}(\TR) &=
(\TR-1)^{\ell} \int_{\infty}^{\TR} \!\! d\TR' \!\! \int_{-1}^{1} \!\! d\notu \; \frac{P_{\ell}(\notu) \tsf^{\FOZO} s^{\FOZO}}{(\TR-1)^{\ell+1}}
\nn \\
&-
\frac{1}{(\TR-1)^{\ell+1}} \int_{1}^{\TR} \!\! d\TR' \!\! \int_{-1}^{1} \!\! d\notu \; (\TR-1)^{\ell} P_{\ell}(\notu) \tsf^{\FOZO} s^{\FOZO}
\nn \\
&-
\frac{(2 \ell+1)}{2}  \int_{-1}^{1} d\notu P_{\ell}(\notu) \; (\tsf^{\FOZO})^{2}\,.
\end{align}
This completes the formal solution for the modes of the trace of the metric
perturbation in the extremal limit in integral form.

The angular integrals in Eq.~\eqref{eq:g2ExtSolution} can be evaluated
in closed form using the Legendre decomposition of the scalar field
and the source function. From Eq.~\eqref{eq:LegendreDecomp} and
Eq.~\eqref{ScalarSourceFunctions} we have
\begin{align} \nn
\tsf^{\FOZO}(\TR,\notu) &= \sum_{k=1}^{\infty} \tsf^{\FOZO}_{k}(\TR) P_{k}(\notu)\\ \nn
s^{\FOZO}(\TR,\notu) &= \sum_{j=1}^{\infty} s^{\FOZO}_{j}(\TR) P_{j}(\notu)\,,
\end{align}
where the sums are over odd integers in both cases. Using the
orthogonality of Legendre functions, the $\ell=0$ mode is given by
\begin{align}\label{eq:g2L0}
\tg_{0}^{\SOZO}(\TR) = \sum_{k=1}^{\infty}\frac{2}{2k+1}\Bigg[ & \int_{\infty}^{\TR} d\TR' \,\frac{ \tsf_{k}^{\FOZO}(\TR') s_{k}^{\FOZO}(\TR')}{\TR'-1}
\nn -\frac{1}{\TR-1}\,\int_{1}^{\TR} d\TR'\,\tsf_{k}^{\FOZO}(\TR') s_{k}^{\FOZO}(\TR')
\nn \\
&- \frac{1}{2}\,\tsf_{k}^{\FOZO}(\TR)^{2} \Bigg] ~.
\end{align}
For general $\ell$, the integration over $\notu$ can be expressed in
terms of the standard $3j$-symbols. The resulting expression is
\begin{align}
\label{eq:g2GeneralL}
 \tg_{\ell}^{\SOZO}(\TR) &=
 \sum_{k,j} 2	\begin{pmatrix}
  \ell & k & j \\
  0 & 0 & 0
\end{pmatrix}^2 \times
\nn \Bigg[(\TR-1)^{\ell}
\int_{\infty}^{\TR}d\TR' \frac{\tsf_{k}^{\FOZO}(\TR') s_{j}^{\FOZO}(\TR')}{(\TR'-1)^{\ell+1}}
\nn \\
&- \frac{1}{(\TR-1)^{\ell+1}}\int_{1}^{\TR}d\TR' (\TR'-1)^{\ell} \,\tsf_{k}^{\FOZO}(\TR') s_{j}^{\FOZO}(\TR')
\nn \\
&- \frac{2\ell+1}{2}\,\tsf_{k}^{\FOZO}(\TR)\,\tsf_{j}^{\FOZO}(\TR) \Bigg]
\end{align}
The radial integrals in Eqs.~\eqref{eq:g2L0}
and~\eqref{eq:g2GeneralL}, though still complicated, are more
tractable than the integrals that result from expressing the source
function in terms of the modes of the scalar field in
Eq.~\eqref{eq:g2-FirstSolution}.

We have not yet obtained a closed-form expression for the trace of the
metric perturbation on the extremal background. The main difficulty,
apparent in Eqs.~\eqref{eq:g2ExtSolution}-\eqref{eq:g2GeneralL}, is
that the source term depends on the full tower of Legendre modes of
the scalar field. Using our expressions for the modes of the scalar
field and its source, Eq.~\eqref{eq:General-Mode} and
Eq.~\eqref{eq:nasty-long-source}, it is possible to evaluate
individual terms in these sums. However, we have not been able to
perform the sums themselves. Indeed, the analytic results for the
individual terms are sufficiently complicated that we turn to
approximations and numerical analysis, which we discuss in the next
section.

\section{Trace of the Metric Perturbation: Properties}
\label{sec:MetricPertProperties}

The results of Sec.~\ref{sec:ScalarFieldProperties} suggest that the
first three or four modes of the scalar field capture most of its
physics, and should be sufficient for analyzing the behavior of the
trace of the metric perturbation. But first, let us consider a few
important properties of the modes $\tg_{\ell}^{\SOZO}$ that can be
extracted from the integral form of the solution.

At large radius, $\TR \gg 1$, the second line of
Eq.~\eqref{eq:g2ExtSolution} dominates and the leading behavior of the
mode is $g_{\ell}^{\SOZO} \sim \TR^{-(\ell+1)}$. This is because
the first line of Eq.~\eqref{eq:g2ExtSolution} decays with a higher
power of $\TR$,
while the third line is proportional to $(\tsf^{\FOZO})^{2}$ and
therefore decays as $\TR^{-2(\ell +1)}$. Near the horizon the first
and third line of Eq.~\eqref{eq:g2ExtSolution} dominate. The
asymptotic behavior as $\TR \to 1$ is most easily extracted by first
evaluating Eq.~\eqref{eq:g2-FirstSolution} at
$\TR = \TR_{+} = 1 + \eps$, changing the integration variable to
$\eta = (\TR'-1)/\eps$, and then taking the $\eps \to 0$ limit, which
gives
\begin{gather}\label{eq:gLhorizon}
  g_{\ell}^{\SOZO}(1) = S_{\ell}^{\SOZO}(1)\,\int_{\infty}^{1} d\eta \, Q_{\ell}(\eta) ~.
\end{gather}
The overall factor of $S_{\ell}^{\SOZO}(1)$, the source function
evaluated at the extremal Kerr horizon, can be expressed in terms of the
source functions for the scalar field. Evaluating
Eq.~\eqref{eq:SLdefinition} at $\TR=1$, using
$\lim_{\TR \to 1} \tD (\partial_{\TR} \tsf)^2 = 0$, and performing the
angular integral yields
\begin{multline}
\label{eq:SLHorizon}
  S_{\ell}^{\SOZO}(1) =  - \sum_{k,j=1}^{\infty} \frac{2(2\ell+1)}{\sqrt{j(j+1)k(k+1)}}\,
  \begin{pmatrix}
 \ell & j & k \\
 0 & 0 & 0
\end{pmatrix}
\begin{pmatrix}
 \ell & j & k \\
 0 & 1 & -1
\end{pmatrix} \,s_{j}^{\FOZO}(1)s_{k}^{\FOZO}(1)  ~, 
\end{multline}
where again the result is expressed in terms of $3j$-symbols.

\begin{figure}[t]
  \centering
    \includegraphics[width=0.9\columnwidth{}]{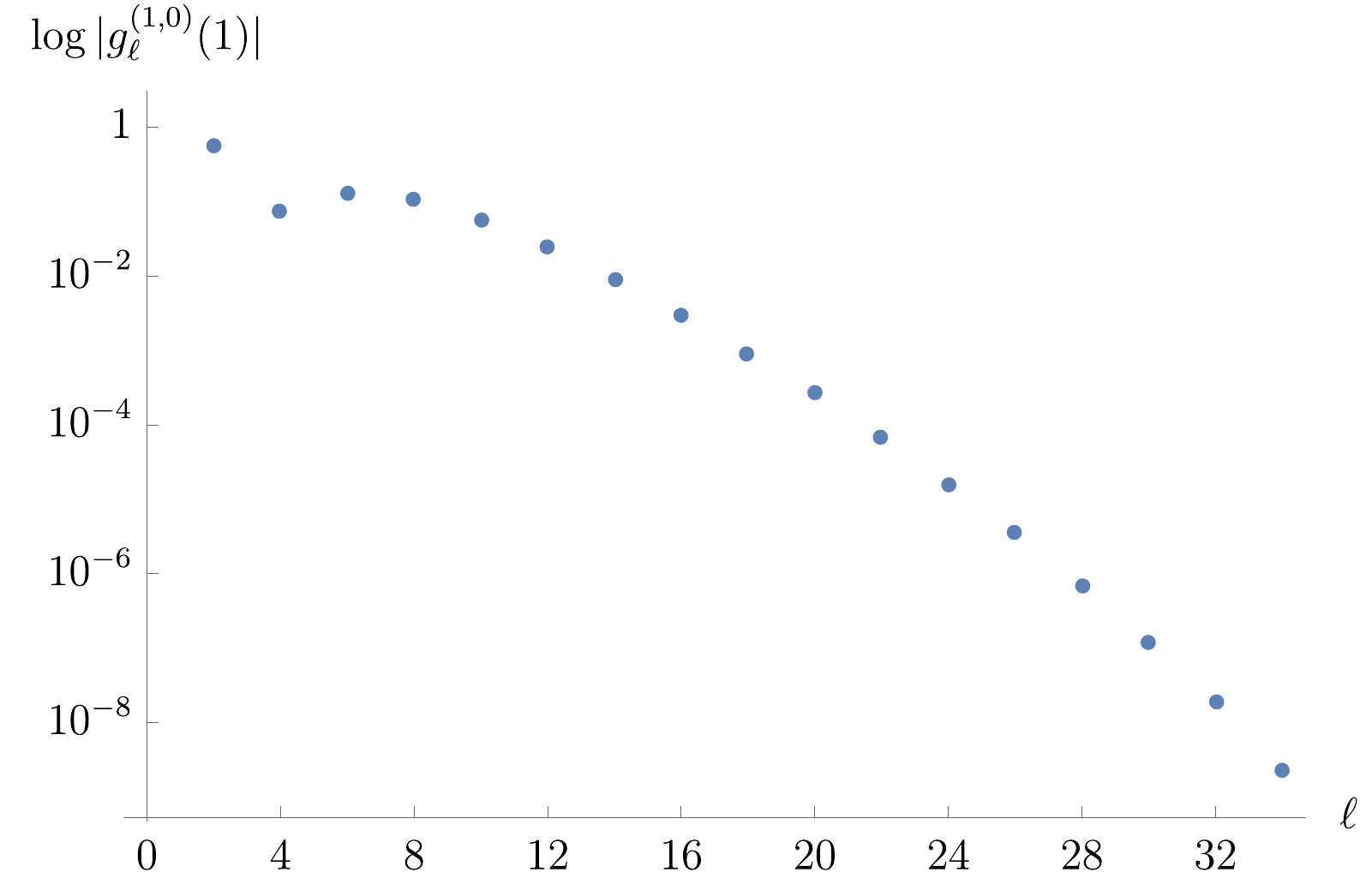}
    \caption{
      The (absolute value of) Legendre modes of the trace of
      the metric perturbation evaluated at the horizon of the extremal
      background, on a logarithmic scale, as a function of harmonic
      number $\ell$. }
  \label{fig:ghorPlot1}
\end{figure}

For $\ell \geq 2$ (recall that $\ell$ is even) the integral in
Eq.~\eqref{eq:gLhorizon} converges: 
\begin{gather}
  g_{\ell}^{\SOZO}(1) =  - \frac{1}{\ell(\ell+1)}\,S_{\ell}^{\SOZO}(1) ~.
\end{gather}
In this case the mode is finite at the horizon of the extremal
background, just like the modes of the scalar field. The values
$g_{\ell}^{\SOZO}(1)$ are plotted against $\ell$ in
Fig.~\ref{fig:ghorPlot1}. Rather than falling off
monotonically with $\ell$, a feature is observed at $\ell=4$. This mode is suppressed relative to
the $\ell=6$ and $\ell=8$ modes.

For the $\ell=0$ mode the integral in Eq.~\eqref{eq:gLhorizon} does
not converge. In this case the mode diverges logarithmically as $\TR \to 1$. Its behavior near the horizon is captured by the first integral in Eq.~\eqref{eq:g2L0}. Expanding the integrand in powers of $(\TR'-1)$ and extracting the log term from the integral gives
\begin{gather}\label{eq:MonopoleModeAtHorizon}
  \lim_{\TR \to 1} g_{0}^{\SOZO}(\TR) \sim S_{0}^{\SOZO}(1)  \log(\TR-1) ~,
\end{gather}
The coefficient $S_{0}^{\SOZO}(1)$ may be evaluated from Eq.~\eqref{eq:SLHorizon}, which for $\ell=0$ reduces to a single sum
\begin{align}\label{eq:LogCoefficient}
  S_{0}^{\SOZO}(1) = &\,\, - \sum_{k=1}^{\infty} \frac{2}{k(k+1)(2k+1)}\,(s_{k}^{\FOZO}(1))^2 \\ \nonumber
     \simeq &\,\, -3.52572
     \,.
\end{align}
This is the same value that is obtained from the ratio of the numerical solution for $g_{0}^{\SOZO}(\TR)$ and $\log(\TR-1)$, evaluated as $\TR \to 1$. As we will discuss at the end of this Section, the log divergence in $g_{0}^{\SOZO}(\TR)$ need not imply the existence of a naked curvature singularity at the perturbed horizon. 

\begin{figure}[t]
  \centering
  \includegraphics[width=0.9\columnwidth{}]{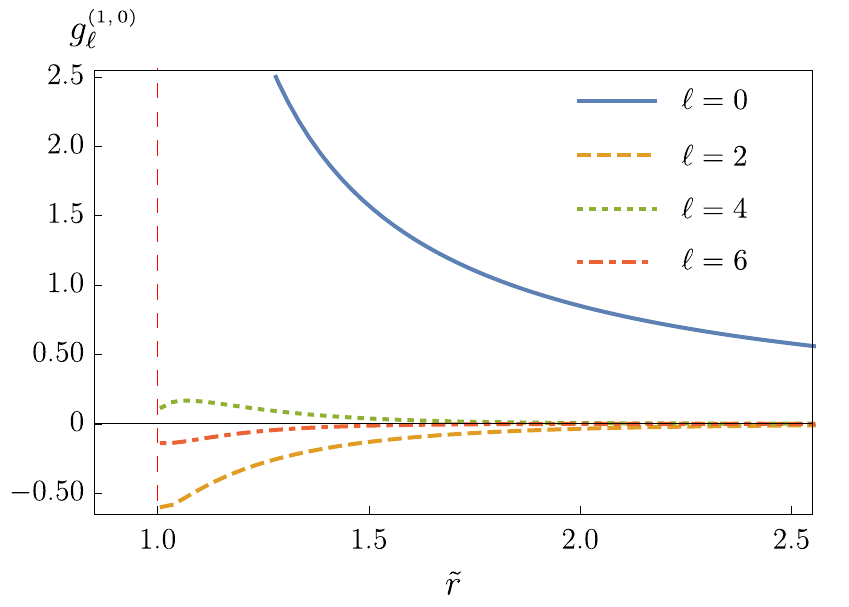}
  \caption{%
    The first four radial modes of the trace of the metric
    perturbation as a function of $\TR$. The dashed vertical line
    indicates the horizon of the extremal background. The $\ell=0$
    mode exhibits a logarithmic divergence as $\TR \to 1$.  }
  \label{fig:gModeFunctions1}
\end{figure}

As explained at the end of the previous Section, analytical results
for the $\TR$-dependence of the modes $\tg_{\ell}^{\SOZO}$ are
sufficiently complicated that a numerical analysis is called for. The
first four modes of the trace of the metric perturbation are shown in
Fig.~\ref{fig:gModeFunctions1}.
These are numerical solutions, obtained with closed form expressions
for the source that include contributions from modes of the scalar
field with $\ell \leq 21$. It is immediately apparent that the
$\ell=0$ mode dominates the trace of the metric perturbation, even
away from the log-divergent behavior near $\TR = 1$.

\begin{figure}[t]
  \centering
  \includegraphics[width=0.9\columnwidth{}]{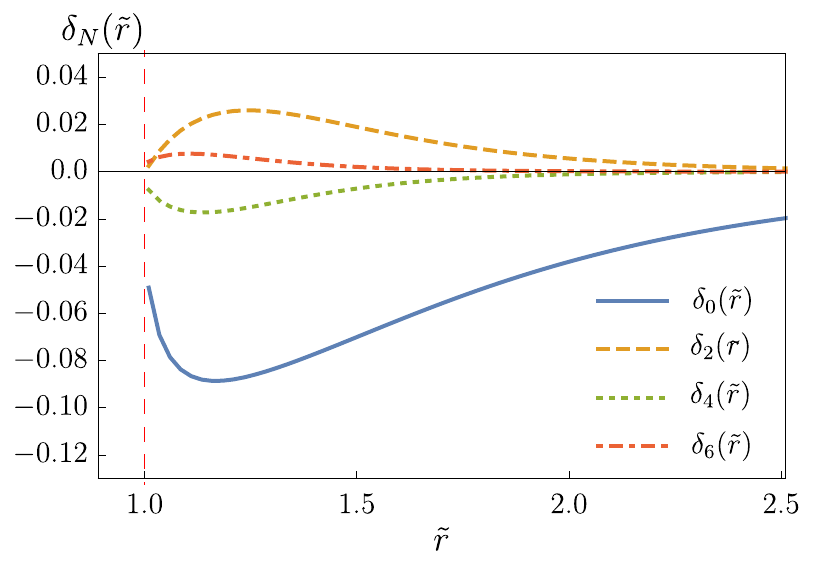}
  \caption{%
    The fractional difference between the trace of the metric
    perturbation at $\theta = 0$, and its approximation including only
    modes with $\ell \leq N$.  }
  \label{fig:deltaPlot1}
\end{figure}

We expect, based on the behavior shown in
Fig.~\ref{fig:gModeFunctions1} that $\tg^{\SOZO}$ is well-approximated
by its first few Legendre modes. In the case of the scalar field, a
similar conclusion was justified by examining mode-by-mode
contributions to the ADM energy. But before applying this norm to the modes of the trace of the metric perturbation, we first consider the
fractional difference between $\tg^{\SOZO}(\TR,\psi)$ evaluated at $\psi = 1$,
and an approximation by its first $N$ modes
\begin{gather}\label{eq:gFracError}
  \delta_{N}(\TR) = 1 - \frac{1}{\tg^{\SOZO}(\TR,1)}\,\left( \sum_{\ell=0}^{N} \tg^{\SOZO}_{\ell}(\TR) \right) ~.
\end{gather}
The fractional difference for $N=0,2,4,6$ is shown in Fig.~\ref{fig:deltaPlot1}.
As expected, the log-divergence of the $\ell=0$ modes means that the
fractional difference $\delta_{0}(\TR) \to 0$ as $\TR \to 1$. 
Figure~\ref{fig:deltaPlot1} shows that if one wishes an accuracy of no
more than about $10\%$, then retaining only the $\ell=0$ mode
suffices.  To obtain a higher accuracy, more modes are needed. In
particular, since the $\ell=4$ mode is suppressed at $\TR=1$ relative
to the $\ell=6$ and $8$ modes, one must include modes up to $\ell=6$
to obtain uniform accuracy of at least one percent.
Observe, however, that if one is interested in the trace of the metric
perturbation outside a larger radius, such as for $\TR \gtrsim 2$, then
$\tg^{\SOZO}$ is approximated at better than percent precision with
only the first two or three modes.

\begin{figure}[t]
  \centering
  \includegraphics[width=0.8\columnwidth{}]{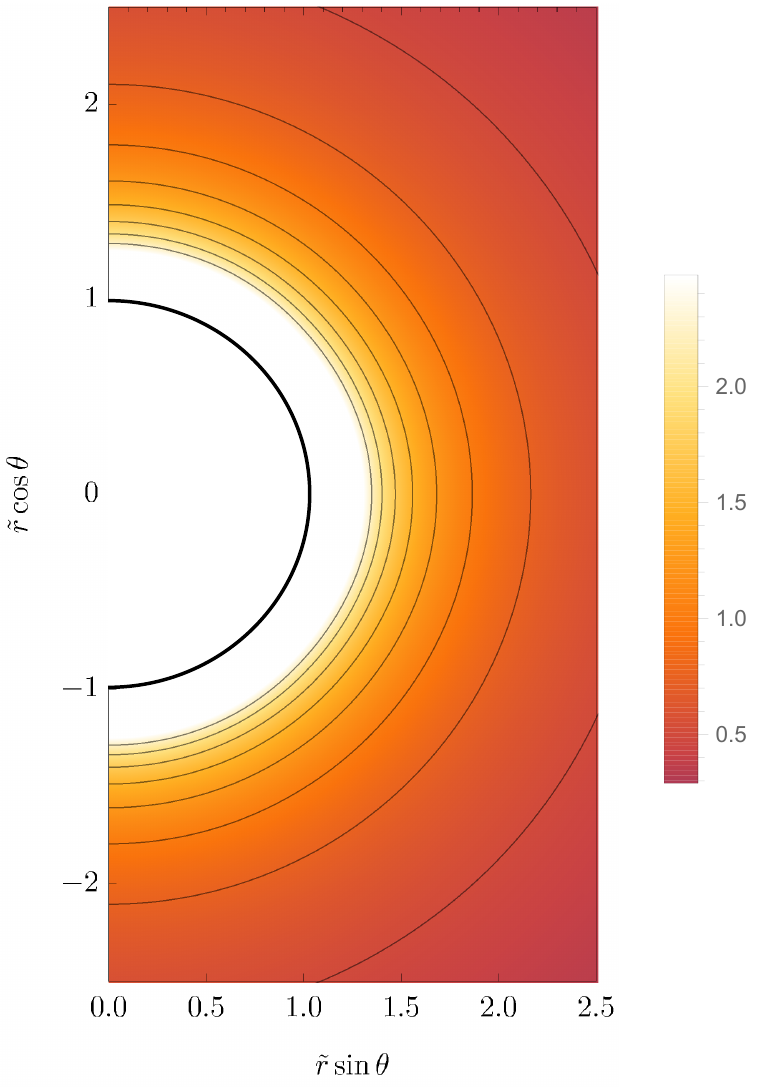}
  \caption{%
    The trace of the metric perturbation on the extremal background,
    approximated by its first four Legendre modes. Near the extremal
    horizon $\TR=1$ (the solid line) the logarithmic divergence of the
    monopole term dominates.  }
  \label{fig:gxyTotPlot}
\end{figure}

We conclude that Legendre modes $\tg^{\SOZO}_{\ell}$ with
$\ell \leq 4$ capture almost all of the physics of $\tg^{\SOZO}$,
except near $\TR=1$ where the $\ell=6$ and $\ell=8$ modes may be
required. Note that the fractional error defined in
Eq.~\eqref{eq:gFracError} puts a bound on the fidelity of our
approximation of $\tg^{\SOZO}$ at $\notu=1$ ($\theta = 0$), where
$P_{\ell}(\notu) = 1$ for all $\ell$. However, since the $\ell=0$ mode
dominates, and $-1 \leq P_{\ell}(\notu) \leq 1$ for $\ell \geq 2$, the
fractional error $\delta_{N}(\TR)$ gives an upper bound on the
fidelity of the approximation in the full $(\TR,\notu)$ plane. An
approximation of $\tg^{\SOZO}$ by its first four Legendre modes is
shown in Fig.~\ref{fig:gxyTotPlot}.

The analysis above uses modes of the scalar field with $\ell \leq 21$
to approximate the source for the trace of the metric
perturbation. However, as we saw in
Sec.~\ref{sec:ScalarFieldProperties}, the first three modes of the
scalar field account for most of its contribution to the ADM
energy. Indeed, the behavior of the first few modes of $\tg^{\SOZO}$
is largely unchanged if we include fewer modes of the scalar field. In
particular, we achieve comparable results for the mode
$\tg^{\SOZO}_{\ell}$ by approximating the scalar field by its first
$N=\ell+1$ modes.

\begin{figure}[tbp]
  \centering
  \includegraphics[height=5.7cm]{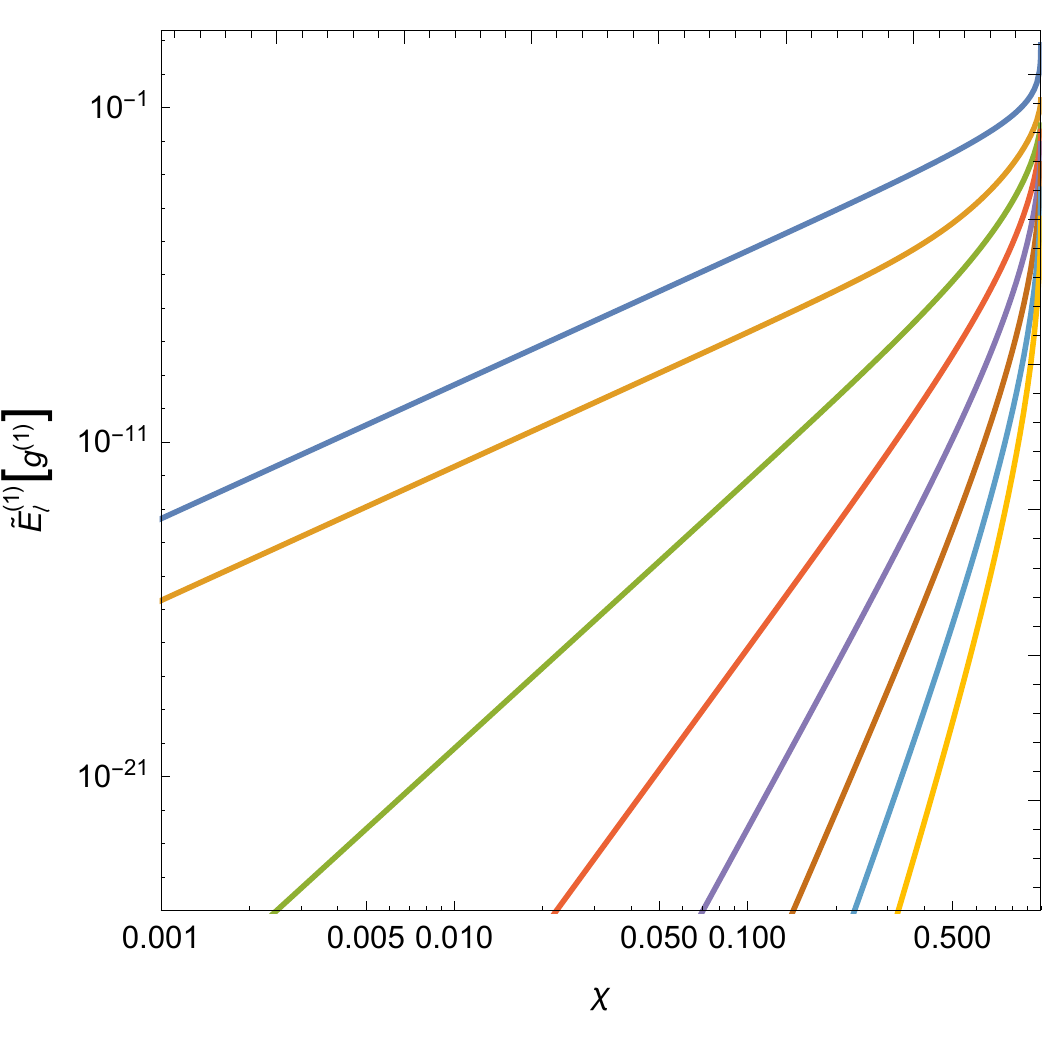}
  \includegraphics[height=5.7cm]{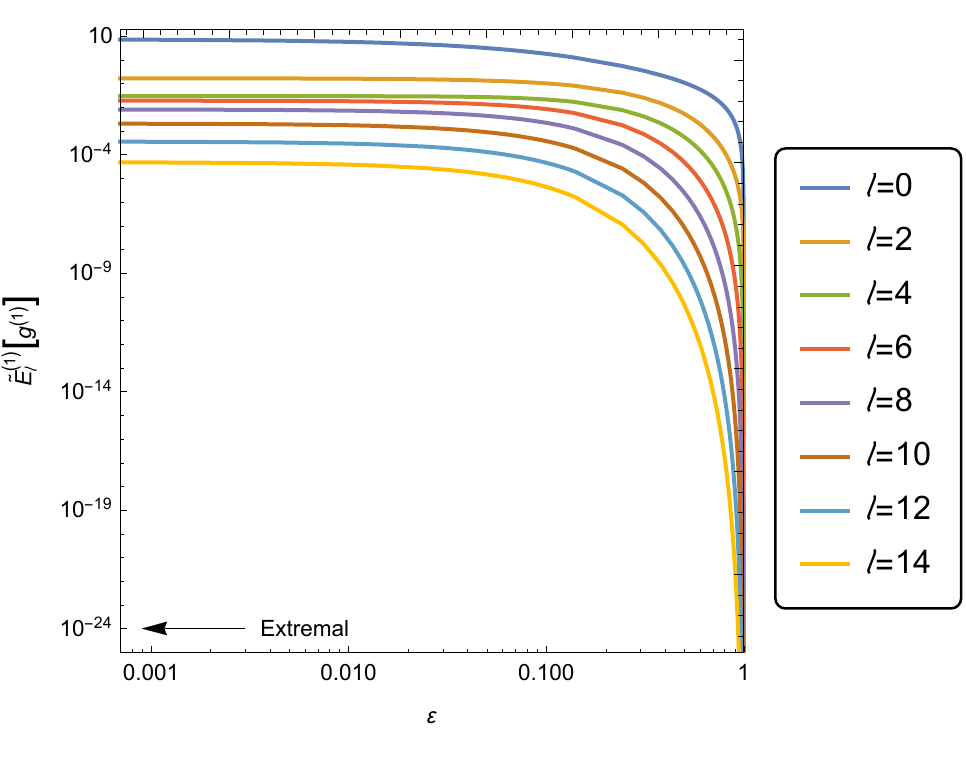}
  \caption{%
Using the scalar ADM energy functional [defined in
Eq.~\eqref{eq:tildeEkSO}] to measure the contributions to the trace of
the metric perturbation $g^{(1)}$ in each $\ell$ mode. Notice that the
modal energy in the extremal limit is regular, despite the logarithmic
divergence in the $\ell=0$ mode. This is emphasized by making the
horizontal axis the extremality parameter $\varepsilon =
\sqrt{1-\chi^{2}}$. Note that each curve becomes horizontal going
toward the left edge of the right panel.
  }
  \label{fig:ElhCSofa}
\end{figure}

Just as in the case of the scalar field, we can study the convergence
of numerical solutions with $\varepsilon > 0$ in the limit as
$\varepsilon \to 0$, using the numerical technique of
Stein~\cite{Stein:2014xba}. We use the ADM energy functional, given in
Eq.~\eqref{eq:tildeEkSO}, as a norm for the modes
$g_{\ell}^{\SO}$. Despite the fact that $g_{0}^{\SO}$ has a
logarithmic divergence at the horizon, this divergence is integrable
in the norm of Eq.~\eqref{eq:tildeEkSO}. This results in each of the
$\ell$ modes' energies being convergent as $\varepsilon\to 0$, as seen
in Fig.~\ref{fig:ElhCSofa}. 

As in Sec.~\ref{sec:ScalarFieldProperties}, we can again plainly see the differences between the trace of the metric perturbation in the slowly rotating and extremal limits, with the higher $\ell$ modes becoming fractionally more important as rotation increases. In the slowly rotating case the monopole mode dominates, with the energy of the modes falling off like $\chi \ll 1$ to a power that is monotonically increasing with $\ell$. For $\chi$ sufficiently close to zero, the monopole and quadrupole modes of the trace of the metric perturbation account for the ADM energy to better than 1 part in $10^{12}$. Near extremality, the monopole mode still dominates, but the contributions to the ADM energy from the first eight modes ($\ell \leq 14$) span a range of only about 8 decades. As pointed out earlier, approximating the trace of the metric perturbation to better than $1\%$ fidelity in the extremal limit requires at least the first four modes ($\ell \leq 6$). Thus, as with the scalar field, modes of the trace of the metric perturbation that are not relevant in the limit of slow rotation become more important near extremality.

Finally, we turn to the question of the significance of the
logarithmic divergence of the $\ell=0$ mode, as seen in
Eq.~\eqref{eq:MonopoleModeAtHorizon}.  Without a full metric tensor
perturbation solution, we cannot determine if this is a true
curvature singularity of the perturbed spacetime.  However, several
pieces of evidence suggest that a divergence in the trace of the metric perturbation need not be
a problem.

First, note that shifts within the Kerr family of solutions
manifest perturbatively as divergences of the trace of the metric perturbation. 
Consider shifting the Kerr metric by $M \to M + \delta M$ and $a \to a + \delta a$, 
and expanding the resulting metric in powers of $\delta M \ll M$ and
$\delta a \ll a$. This is made formal by the coefficients in the
Taylor series, which are given by
\begin{align}
  \label{eq:pert-params-g-defs}
  (\delta_{M} g)_{ab} &\equiv \frac{\pd}{\pd M} g_{ab}
  \,, &
  (\delta_{a} g)_{ab} &\equiv \frac{\pd}{\pd a} g_{ab}
   \,,
\end{align}
and so on for higher derivatives, e.g.~$\delta_{M,M} g_{ab}$.
The perturbed shifted Kerr metric tensor satisfies the perturbative Einstein equations
order-by-order, since they are simply shifts of the Kerr parameters $M$ and $a$.
A straightforward calculation in Boyer-Lindquist coordinates, however, shows that
\begin{align}
  \label{eq:pert-params-g-tr}
  \Tr[\delta_{M}g] &= 0
  \,, &
  \Tr[\delta_{a}g] &= \frac{4 a \cos^{2} \theta }{ \Sigma }
  \,, \\
  \Tr[\delta_{M,M}g] &= \frac{8 r^{2} }{ \Delta^{2} }
  \,, &
  \Tr[\delta_{a,M}g] &= \frac{ p_{1}(r,\cos\theta) }{ \Delta^{2}\Sigma }
  \,, &
  \Tr[\delta_{a,a}g] &= \frac{ p_{2}(r,\cos\theta) }{ \Delta^{2}\Sigma^{2} }
  \,,
\end{align}
where $p_{1,2}$ are certain regular polynomials of $r$ and
$\cos\theta$, neither of which have zeros at the roots of either $\Delta$
or $\Sigma$.  Although the Kerr spacetime is regular, as are the linearly or
quadratically shifted spacetimes, we immediately see that the
traces of the above metric perturbations diverge at either the event
and Cauchy horizons where $\Delta=0$, or the (ring) curvature
singularity where $\Sigma=0$, or both.  Only the blowup at the
curvature singularity can be seen as physically significant.  When
$a\to M$, in Boyer-Lindquist coordinates, the curvature singularity
appears on the horizon, at the equator. 

Next we can bring in arguments that are more specific to the dCS
problem.  As noted by Stein~\cite{Stein:2014xba}, the (inner)
Cauchy-horizon divergence of the scalar found by Konno and
Takahashi~\cite{Konno:2014qua} may result in a divergence of the trace
of the metric perturbation at the inner horizon; however, this
divergence would be hidden by the outer horizon for all $|a|<M$.
Indeed, Ref.~\cite{Stein:2014xba}, found that the trace of the
metric perturbation is regular in the exterior for all $|a|<M$.
However, in the limit $a\to M$, the inner and outer horizons
confluence.
Further, in the slow-rotation expansion, the horizon was seen to be
shifted outward to
$r_{\text{hor}} = r_{\text{hor,Kerr}} + (915/28672) \zeta M \chi^{2} +
{\cal{O}}(\chi^{4})$~\cite{Yagi:2012ya}.
Thus, one may expect that in the extremal limit, the perturbed horizon
may again be exterior to that of the background, hiding any potential
singularity.  Finally, it may be the case that the extremality
condition in dCS is shifted away from $J=M^{2}$.

All of these arguments suggest that the horizon singularity in
$g_{0}^{\SOZO}$ may not be a physical curvature singularity, but
rather a coordinate singularity.  A full tensor perturbation solution
is needed to confirm this conjecture.

\section{Discussion}
\label{sec:Conclusions}

This paper explored rotating black holes in dCS. Using an effective
field theory treatment of dCS, we worked in the decoupling limit where
dCS corrections are small perturbations from GR solutions. We have
further focused on BHs that spin at the maximal Kerr rate, the
so-called extremal limit. With these assumptions, we then
solved for the dynamical scalar field in closed analytic form, obtaining
a Legendre decomposition that is dominated by its dipole and octupole
terms. The radial structure of this decomposition includes 
natural logarithms and arctangents, unlike the simple polynomial
results obtained in the slow-rotation limit. We then solved for the
Legendre decomposition of the trace of the metric perturbation. 
Retaining Legendre modes with $\ell \leq 6$ is sufficient to ensure a 
fidelity of at least $99\%$ relative to numerical solutions for both the
scalar field and the trace of the metric perturbation.

The trace of the metric perturbation in the Lorenz-like gauge exhibits a
logarithmic divergence at the location of the background (extremal
Kerr) event horizon.  Several pieces of evidence, discussed in
Sec.~\ref{sec:MetricPertProperties}, show that it need not be a
physical curvature singularity.  We conjecture that it is only a
coordinate singularity, but a full metric tensor perturbation solution
is needed to confirm this conjecture.

The techniques we employed rely heavily on Legendre expansions, but
our work suggests that these can be truncated at a finite mode number
without losing much of the overall behavior of the fields. In
particular, if one wishes to carry out astrophysical tests of GR,
certain observables may be sensitive to only certain regions of
spacetime that need not include the horizon. For example, BH shadow
observations are most sensitive to the location of the light-ring,
while astrophysical observations of the energy spectrum of radiation
emitted by accretion disks are most sensitive to the location of the
innermost stable circular orbit. For such observations, it may suffice
to keep only the first few modes in a Legendre expansion provided the
BH is not rotating maximally.
The reason here is two-fold.  First, the light-ring and ISCO are both
pushed away from the horizon as the spin decreases, and the
approximation by a finite number of Legendre modes improves away from
the immediate vicinity of the horizon.  Second, our studies
indicate that, in general, the fidelity of an approximation with
a fixed number of modes improves away from extremality, as shown
in~\cite{Stein:2014xba}.

The results obtained here open the door for new investigations of
rotating BHs in dCS gravity. As mentioned in the introduction, 
the methods we employed can be applied to the next-order term in a 
near-extremal expansion. They also generalize to BHs that rotate with arbitrary spins, and 
we have already obtained closed-form results in that case for the first few modes of the scalar field (and numerical
results for all other modes). Ultimately, of course, our goal is to
go beyond the trace and determine the full metric perturbation of dCS BHs. Previous work on this problem was limited by 
incomplete results for the scalar field, but with the results in this paper we are now in a position to attack the full problem.

\section*{Acknowledgments}

We would like to thank the Kavli Institute for Theoretical Physics for
their hospitality during the completion of this work.
We would also like to thank Kent Yagi, Frans Pretorius, and Albion Lawrence for useful discussions.
RM acknowledges support from a Loyola University Chicago Summer
Research Stipend.
LCS~acknowledges that support for this work was provided by NASA
through Einstein Postdoctoral Fellowship Award Number PF2-130101
issued by the Chandra X-ray Observatory Center, which is operated by
the Smithsonian Astrophysical Observatory for and on behalf of the
NASA under contract NAS8-03060, and further acknowledges support from
the NSF grant PHY-1404569.
NY acknowledges support from NSF CAREER Award PHY-1250636. The
research of RM and NY was supported in part by the National Science
Foundation under Grant No.~NSF PHY11-25915.
Some calculations used the computer algebra-system \textsc{Maple}, in
combination with the \textsc{GRTensorII} package~\cite{grtensor}.
Other calculations used the computer algebra-system
\textsc{Mathematica}, in combination with the \textsc{xTensor}
package~\cite{Martin-Garcia:2007bqa, MartinGarcia:2008qz,
  Brizuela:2008ra}.
Finally, RM (\href{https://twitter.com/mcnees}{@mcnees}) and LCS
(\href{https://twitter.com/duetosymmetry}{@duetosymmetry}) would like
to thank Twitter for facilitating discussion during the early stages
of this collaboration.

\appendix
\section{Solution of the Scalar Equation of Motion}
\label{app:GeneralSolution}
The equations of motion for the scalar field and the trace of the
metric perturbation have the same general form, so let us briefly
establish some conventions for the solutions of such equations. First,
consider an equation of the form
\begin{gather}\label{eq:InhomogeneousEqn}
  \partial_{\TR}(\tD \partial_{\TR} I_{\ell}) - \ell(\ell+1)\,I_{\ell} = K_{\ell}
\end{gather}
with some source $K_{\ell}$. Denote by $H_{\ell}^{+}$ and
$H_{\ell}^{-}$ the solutions of the homogeneous equation
\begin{gather} \label{eq:HomogeneousEqn}
  \partial_{\TR}(\tD \partial_{\TR} H_{\ell}^{\pm}) - \ell(\ell+1)\,H_{\ell}^{\pm} = 0~,
\end{gather}
with $H_{\ell}^{+}$ regular at $\TR_{+}$ and $H_{\ell}^{-} \to 0$ at
$\TR \to \infty$.
We use the method of variation of parameters to find the general
solution to the inhomogeneous equation.
The solution of~\eqref{eq:InhomogeneousEqn}
that is both regular at $\TR_{+}$ and goes to zero as $\TR \to \infty$
can be expressed in terms of the homogeneous solutions and the source
as
\begin{multline}\label{eq:GeneralScalarEOMSolution}
  I_{\ell} = \frac{1}{W_{\ell}} \times \left( H_{\ell}^{+}(\TR) \int_{\infty}^{\TR} d\TR' H_{\ell}^{-}(\TR') K_{\ell}(\TR')
  - H_{\ell}^{-}(\TR) \int_{\TR_{+}}^{\TR} d\TR' H_{\ell}^{+}(\TR') K_{\ell}(\TR')\right) ~.
\end{multline}
Here a factor of $\tD$ has canceled inside each integral, allowing us
to pull out a constant $W_{\ell}$; this constant depends on the Wronskian
of the homogeneous solutions,
\begin{align} \label{Wronskian}
  W_{\ell} \equiv &\,\,\tD \times W[H_{\ell}^{-}, H_{\ell}^{+}] \\ \nonumber
  = &\,\, \tD \times \left(H_{\ell}^{-} \partial_{\TR} H_{\ell}^{+} - H_{\ell}^{+} \partial_{\TR} H_{\ell}^{-} \right) ~.
\end{align}
It is straightforward to verify that this is constant using
Eq.~\eqref{eq:HomogeneousEqn}.

For the Kerr background, the homogeneous solutions can be written as
\begin{align} \label{eq:Hplus}
  H_{\ell}^{+}(\TR) = &\,\, c_{\ell}^{+}\, (1-\chi^2)^{\tfrac{\ell}{2}}\, P_{\ell}\left(\frac{\TR-1}{\sqrt{1-\chi^2}}\right)\\ \label{eq:Hminus}
  H_{\ell}^{-}(\TR) = &\,\, c_{\ell}^{-}\, (1-\chi^2)^{-\tfrac{\ell+1}{2}}\, Q_{\ell}\left(\frac{\TR-1}{\sqrt{1-\chi^2}}\right) ~.
\end{align}
where $P_{\ell}(\cdot)$ and $Q_{\ell}(\cdot)$ are Legendre functions
of the first and second kind, respectively.  A standard identity for
Legendre functions then gives the factor
$W_{\ell}=c_{\ell}^{+}\,c_{\ell}^{-}$.

The factors of $\sqrt{1-\chi^2}$ in
Eq.~\eqref{eq:Hplus}-\eqref{eq:Hminus} have been chosen so that the
extremal limit, $\chi \to \pm 1$, is regular. Otherwise, the overall
normalization factors $c_{\ell}^{\pm}$ are arbitrary. A convenient
choice is to set
\begin{gather}\label{eq:HomSolutionNormalization}
  c_{\ell}^{+} = \frac{\ell!}{(2\ell-1)!!} \,,
  \qquad
  c_{\ell}^{-} = \frac{(2\ell+1)!!}{\ell!} \,.
\end{gather}
Then $W_\ell = 2\ell + 1$, and in the extremal limit
the homogeneous solutions are simply
\begin{align} \label{eq:HplusExt}
  \lim_{|\chi| \to 1} H_{\ell}^{+} = &\,\, (\TR -1)^{\ell} \\ \label{eq:HminusExt}
  \lim_{|\chi| \to 1} H_{\ell}^{-} = &\,\, \frac{1}{(\TR -1)^{\ell+1}} ~.
\end{align}
We adopt this normalization throughout
Secs.~\ref{sec:ScalarFieldSolutions} and
\ref{sec:TraceMetPertSolution}.

\section{Expressions for Radial Modes}
\label{app:higher-modes}

The radial mode function for general $\ell$ is given in
Eq.~\eqref{eq:General-Mode}. In this form, each mode depends on
$2\ell+1$ coefficients $\alpha_{\ell,k}$ and $\beta_{\ell,k}$. The
coefficients for the modes up to $\ell=9$ are given below.

\begin{center}
\bgroup
\def\arraystretch{1.5}
\setlength\tabcolsep{0.5em}
\begin{tabular}{c|cc|cc|cc|cc|cc}
   k &  $\alpha_{1,k}$ &  $\beta_{1,k}$ &  $\alpha_{3,k}$ &  $\beta_{3,k}$ &  $\alpha_{5,k}$ &  $\beta_{5,k}$ &  $\alpha_{7,k}$ &  $\beta_{7,k}$ & $\alpha_{9,k}$ & $\beta_{9,k}$ \\ \hline
  $0$ & $0$ & $-3$ & $\frac{190}{3}$ & $\frac{15}{2}$ & $-\frac{68}{5}$ & $-\frac{85}{8}$ & $\frac{31719}{70}$ & $\frac{189}{16}$  & $\frac{49019}{1260}$ & $-\frac{1377}{128}$ \\ \hline
  1 & - & $3$ & $-55$ & $-75$ & $\frac{1015}{8}$ & $\frac{2955}{8}$  & $-\frac{22317}{80}$ & $-\frac{4347}{4}$ & $\frac{351417}{896}$ & $\frac{315405}{128}$ \\ \hline
  2 & - & - & $15$ & $75$ & $-\frac{8645}{8}$ & $-\frac{735}{2}$ & $\frac{90819}{20}$ & $\frac{8127}{8}$  & $-\frac{13832005}{896}$ & $-\frac{34155}{16}$\\ \hline
  3 & - & - & - & $-25$ & $\frac{4263}{8}$ & $\frac{4935}{4}$ & $-\frac{44079}{16}$ & $-\frac{33705}{4}$ & $\frac{3351161}{384}$ & $\frac{1065405}{32}$\\ \hline
  4 & - & - & - & - & $-\frac{819}{8}$ & $-\frac{4935}{8}$ & $\frac{21009}{2}$ & $\frac{73395}{16}$ & $-\frac{9069467}{128}$ & $-\frac{1183545}{64}$ \\ \hline
  5 & - & - & - & - & - & $\frac{987}{8}$ & $-\frac{14067}{4}$ & $-\frac{44793}{4}$  & $\frac{3456585}{128}$ & $\frac{6416487}{64}$\\ \hline
  6 & - & - & - & - & - & - & $\frac{1989}{4}$ & $\frac{14931}{4}$ & $-\frac{30089455}{384}$ & $-\frac{292215}{8}$\\ \hline
  7 & - & - & - & - & - & - & - & $-\frac{2133}{4}$  & $\frac{2523125}{128}$& $\frac{2579445}{32}$ \\ \hline
  8 & - & - & - & - & - & - & - & -  & $-\frac{279565}{128}$& $-\frac{2579445}{128}$\\ \hline
  9 & - & - & - & - & - & - & - & -  & - & $\frac{286605}{128}$ \\
\end{tabular}
\egroup
\end{center}

\section{Representations of the Scalar Field}
\label{app:other-scal-field}

Equation~\eqref{eq:General-Mode} provides one representation of the
solution to Eq.~\eqref{ScalarEOM3} for arbitrary harmonic number
$\ell$ after a Legendre decomposition and in an expansion to leading
order in $\zeta$ (i.e.~in the GR deformation) and in $\eps$ (i.e.,
in the extremal limit). This form of the solution depends on $2 \ell + 1$ coefficients
($\alpha_{\ell,k}$ and $\beta_{\ell,k}$) that are fixed by imposing
appropriate boundary conditions on the mode.

In this appendix, we present two additional representations of the
solution to the scalar field evolution equation that may be preferable
in some applications. As in the case of Eq.~\eqref{eq:General-Mode},
these representations will have both advantages and disadvantages that
we will describe in detail. The solutions start by representing the
source function in the extremal limit in terms of a series:
\begin{gather}\label{SeriesRep2}
  s_{\ell}^{\FOZO}(\TR) = \sum_{n=0}^{\infty} \alpha_{\ell,n}\,\frac{2\ell+1}{\TR^{\ell+4+2n}}~,
\end{gather}
where we have introduced the constants
\begin{gather}
  \alpha_{\ell,n} = (-1)^{\frac{\ell-1}{2}} (-1)^{n} \frac{(\ell+2n+2)\, \Gamma(\ell+4+2n)\,\Gamma(\tfrac{1}{2})}{2^{\ell+1+2n} \,\Gamma(n+1) \,\Gamma(\ell+n+\tfrac{3}{2})} ~,
\end{gather}
in terms of the Gamma function $\Gamma(\cdot)$. The factor of $2\ell+1$
in Eq.~\eqref{SeriesRep2} has been introduced to simplify some
expressions, by canceling a similar factor in the denominator of
Eq.~\eqref{eq:extremal-theta}. From here on, different representations
take different routes to arrive at a solution to
Eq.~\eqref{ScalarEOM3} in the extremal limit, so we tackle each of
them separately below.

\subsection{Incomplete Beta Function Representation}

Introducing expansion Eq.~\eqref{SeriesRep2} for the scalar source
into the solution Eq.~\eqref{eq:extremal-theta} for the scalar field:
\begin{multline}\label{sf1}
  \tsf_{\ell}^{\FOZO}(\TR) = \sum_{n=0}^{\infty} \alpha_{\ell,n} \Bigg[(\TR-1)^{\ell} \int_{\infty}^{\TR} \frac{d\TR'}{(\TR'-1)^{\ell+1}\,\TR'^{\ell+4+2n}} \\
  - \frac{1}{(\TR-1)^{\ell+1}}\,\int_{1}^{\TR} d\TR' (\TR'-1)^{\ell}\,\frac{1}{\TR'^{\ell+4+2n}} \Bigg]\,,
\end{multline}
where we have already imposed appropriate boundary conditions. The
integrals can be evaluated in closed-form to obtain
\begin{equation}
\tsf_{\ell}^{\FOZO}(\TR) = \beta_{1}(\TR) + \beta_{2}(\TR) + \beta_{3}(\TR)
\end{equation}
where we have defined
\begingroup
\allowdisplaybreaks[1]
\begin{align}\label{beta-sol:beta1}
  \beta_{1}(\TR) &= -\sum_{n=0}^{\infty} \alpha_{\ell,n} (\TR-1)^{\ell} B_{1/\TR}\bigg(2\ell + 4 + 2n, -\ell \bigg)\,,
  \\ \label{beta-sol:beta2}
  \beta_{2}(\TR) &= \sum_{n=0}^{\infty} \frac{\alpha_{\ell,n} }{(\TR-1)^{\ell+1}} B_{1/\TR}\bigg(2n+3,\ell+1 \bigg) \,,
  \\
  \beta_{3}(\TR) &= - \sum_{n=0}^{\infty} \frac{ \alpha_{\ell,n}}{(\TR-1)^{\ell+1}}\,\frac{\Gamma(\ell+1)\,\Gamma(2n+3)}{\Gamma(\ell+4+2n)}\,,
  \label{beta-sol}
\end{align}%
\endgroup
in terms of the incomplete Beta function
$B_{x}(a,b)$ (see e.g.~Sec.~8.17 of~\cite{NIST:DLMF}),
\begin{equation}
  \label{eq:Beta-func-def}
  B_{x}(a,b) \equiv \int_{0}^{x} t^{a-1}(1-t)^{b-1} dt \,.
\end{equation}
Evaluating at $x=1$ gives the ordinary Beta function,
$B(a,b)=B_{1}(a,b)$.

\bw The sums over $n$ can be evaluated in closed form for $\beta_{2}$
and $\beta_{3}$. The latter can be summed into
\begin{multline}
\beta_{3}(\TR) = \frac{(-1)^{\frac{\ell+1}{2}} \; \ell \; \Gamma(\ell+1) \Gamma(\tfrac{1}{2})}{2^{\ell+2} \Gamma(\ell+ \tfrac{3}{2})}
\frac{1}{(\TR-1)^{\ell+1}}
\Bigg[ \; \ell\,(2\ell+1) \\
- 2 (\ell-1)(\ell+1) {}_{2}F_{1} \big(-\tfrac{1}{2},1;\ell+\tfrac{3}{2};-1\big) \Bigg] ~\,,
\end{multline}
which gives the leading behavior of $\tsf_{\ell}^{(1/2,0)}$ at large
$\TR$. The sum over $n$ for $\beta_{2}$ can be evaluated using the
series representation of the incomplete Beta function appropriate for
Eq.~\eqref{beta-sol:beta2},
\begin{align}\label{beta-series}
B_{x}\big(m,n \big) &= \sum_{j=0}^{n-1}
\frac{(-1)^{j} }{m +j}\,
\frac{\Gamma(n)}{\Gamma(j+1)\,\Gamma(n-j)}
x^{m+j} ~.
\end{align}
Permuting the order of the sums over $j$ and $n$ yields
\begin{align}
\beta_{2}(\TR) &= \frac{(-1)^{(\ell+3)/2} \Gamma(\ell+1) \Gamma(1/2) \Gamma(4+\ell)}{2^{2+\ell}  \Gamma(5/2 + \ell) (\TR-1)^{\ell+1} } \sum_{j=0}^{\ell}
 \frac{(-1)^{j}}{(3+j)(5+j) } \frac{1}{\Gamma(1+j) \Gamma(1 + \ell - j)}
\frac{1}{\TR^{5+j}}
\nn \\
& \times \left[
(5 + j) (2 + \ell) (3 + 2 \ell) \; \TR^{2} \;
{}_{3}F_{2} \left(\frac{3+j}{2},\frac{4+\ell}{2}, \frac{5+\ell}{2}; \frac{5+j}{2}, \frac{3+2\ell}{2};-\frac{1}{\TR^{2}} \right)
\right.
\nn \\
&\left. -
(3+j) (4 + \ell) (5 + \ell)  \;
{}_{3}F_{2} \left(\frac{5+j}{2},\frac{6+\ell}{2}, \frac{7+\ell}{2}; \frac{7+j}{2}, \frac{5+2\ell}{2};-\frac{1}{\TR^{2}} \right)
 \right]\,,
\end{align}
where ${}_{P}F_{Q}(\cdot;\cdot;\cdot)$ is the generalized
hypergeometric function. We have not succeeded in finding a
closed-form expression for the above sum over $j$, but the sum can be
performed explicitly given any value of $\ell$.

One is then only left with the sum over $n$ for $\beta_{1}$. To obtain
an expression for this sum, we start with the following representation
of the incomplete Beta function relevant for
Eq.~\eqref{beta-sol:beta1}:
\begin{align}
B_{\frac{1}{\TR}}\left(2\ell+4+2 n,-\ell\right) &=
\frac{(-1)^{\ell+1} (\ell+4+2 n) \Gamma(2 \ell+4+2n)}{\Gamma(\ell+1)\Gamma(\ell+5+2n)}
\times \left[ \ln \left(\frac{\TR-1}{\TR}\right) \right. \nn \\ & \left. + \frac{1}
{\TR-1} \left(1 - \sum_{k=1}^{\ell+3+2n} \frac{1}{k (k+1)} \frac{1}{\TR^{k}}\right)\right]
\\
&- \frac{\Gamma(2 \ell+4+2n)}{\Gamma(\ell+1)} \frac{1}{\TR^{\ell+3+2n}} \sum_{k=0}^{\ell-2}
(\TR-1)^{k-\ell} \frac{(-1)^{k+1} \Gamma(\ell-k)}{\Gamma(2 \ell+4+2n - k)}\,. \nn
\label{eq:new-beta}
\end{align}
This allows us to write
$\beta_{1}(\TR) := \beta_{4}(\TR) + \beta_{5}(\TR)$, where we have
defined
\begin{multline}
\beta_{4}(\TR) = - (\TR-1)^{\ell} \sum_{n=0}^{\infty} \alpha_{\ell,n} \frac{(-1)^{\ell+1} (\ell+4+2 n) \Gamma(2 \ell+4+2n)}{\Gamma(\ell+1)\Gamma(\ell+5+2n)}
\times \\ \left[ \ln \left(\frac{\TR-1}{\TR}\right) + \frac{1}{\TR-1} \left(1 - \sum_{k=1}^{\ell+3+2n} \frac{1}{k (k+1)} \frac{1}{\TR^{k}}\right)\right]
\end{multline}
\begin{align}
\beta_{5}(\TR) &= -  \frac{1}{\Gamma(\ell+1)} \frac{1}{\TR^{\ell+3}}  \sum_{n=0}^{\infty} \alpha_{\ell,n}
\frac{\Gamma(2 \ell+4+2n)}{\TR^{2n}}
\sum_{k=0}^{\ell-2}
 \frac{(-1)^{k} \;  \Gamma(\ell-k)}{\Gamma(2 \ell+4+2n - k)} (\TR-1)^{k}\,.
\end{align}
The function $\beta_{5}(\TR)$ can be simplified further by performing
the sums to obtain
\begin{align}
\beta_{5}(\TR) &=
(-1)^{(\ell+1)/2} 2^{1+\ell} (1 + \ell) (2 + \ell) \Gamma(4+\ell)
\frac{1}{\TR^{5+\ell}}
\sum_{k=0}^{\ell-2}
(-1)^{k} \frac{\Gamma(\ell-k)}{\Gamma(6-k+2 \ell)}
(\TR-1)^{k}
\nn \\
&\times \left[
\left(k-2 \ell-5\right) \left(3 + 2 \ell\right) \left(k - 4 - 2 \ell\right) \; \TR^{2} \;
{}_{4}F_{3} \left(
  \begin{matrix}
    \frac{4+\ell}{2},\frac{5+\ell}{2},2 + \ell,\frac{5+2\ell}{2}\\
    \frac{3+2\ell}{2},\frac{4-k+2\ell}{2}, \frac{5-k+2\ell}{2}
  \end{matrix}
;-\frac{1}{\TR^{2}} \right)
\right. \nn \\
&\left.
- 2 \left(4 + \ell\right) \left(5 + \ell\right) \left(5 + 2 \ell \right) \;
{}_{4}F_{3} \left(
  \begin{matrix}
    \frac{6+\ell}{2},\frac{7+\ell}{2},3 + \ell,\frac{7+2\ell}{2}\\
    \frac{5+2\ell}{2},\frac{6-k+2\ell}{2}, \frac{7-k+2\ell}{2}
  \end{matrix}
;
-\frac{1}{\TR^{2}} \right)
\right]
\,.
\end{align}

The function $\beta_{4}(\TR)$ can also be simplified by performing
some of the sums in closed form to obtain
\begin{align}
\beta_{4}(\TR) ={}& \frac{(-1)^{\ell}}{\Gamma(\ell+1)}(\TR-1)^{\ell}   \sum_{n=0}^{\infty} \alpha_{\ell,n}  \frac{(\ell+4+2 n) \Gamma(2 \ell+4+2n)}{\Gamma(\ell+5+2n)}
\left[ \ln \left(\frac{\TR-1}{\TR}\right)\right.\nn\\
& \quad\left.{}+\frac{1}
{\TR-1} \left(1 - \sum_{k=1}^{\ell+3+2n} \frac{1}{k (k+1)} \frac{1}{\TR^{k}}\right)\right]
\nn \\
={}& \frac{(-1)^{(3\ell+1)/2}}{2} \frac{\Gamma(\ell+3)}{\Gamma(\ell+1)} (\TR-1)^{\ell-1} \left[ 1 + (\TR-1)  \ln \left(\frac{\TR-1}{\TR}\right) \right]
\nn\\
&\quad{}+ \frac{(-1)^{\ell+1}}{\Gamma(\ell+1)}(\TR-1)^{\ell-1}   \sum_{n=0}^{\infty} \sum_{k=1}^{\ell+3+2n} \gamma_{\ell,n}    \frac{1}{k (k+1)} \frac{1}{\TR^{k}}\,,
\label{eq:beta4}
\end{align}
where we have defined
\begin{equation}
\gamma_{\ell,n} := \alpha_{\ell,n}  \frac{(\ell+4+2 n) \Gamma(2 \ell+4+2n)}{\Gamma(\ell+5+2n)}\,.
\end{equation}
The last term in Eq.~\eqref{eq:beta4} can also be represented as
follows:
\begin{multline}
\sum_{n=0}^{\infty} \sum_{k=1}^{\ell+3+2n} \gamma_{\ell,n}    \frac{1}{k (k+1)} \frac{1}{\TR^{k}}  =
\frac{(-1)^{(\ell+1)/2}}{2} \Gamma(\ell+3) \;
G\left(\frac{1}{\TR},\ell+3\right) \;
\times\\
\times
\sum_{j=0}^{\infty}
\left(\sum_{n=j+1}^{\infty} \gamma_{\ell,n}\right)
\left[\frac{1}{(\ell+4+2j)(\ell+5+2j)} \frac{1}{\TR^{\ell+4+2j}} +
\right. \\
\left. + \frac{1}{(\ell+5+2j)(\ell+6+2j)} \frac{1}{\TR^{\ell+5+2j}} \right]\,,
\end{multline}
\ew
where we have defined the new function
\begin{align}
G(x,N) &:= \sum_{k=1}^{N} \frac{x^{k}}{k (k+1)}\,,
\end{align}
for some $x \in \Re$ and $N \in \mathbb{N}$.  This function is the
first $N$ terms of the Taylor series for $1-\log(1-x)+
x^{-1}\log(1-x)$ about $x=0$.  Notice that the sum in this new
function is finite, and thus $G(1/\TR,N)$ is simply a polynomial in
$1/\TR$. Given a particular value of $\ell$, the remaining sum over
$j$ can be performed explicitly.

\subsection{Radial Series Representation}

Instead of using variation of parameters to solve
Eq.~\eqref{ScalarEOM3}, we will search for a series solution. We thus
insert the ansatz
\begin{gather}\label{eq:SeriesSolution}
  \tsf_{\ell}^{(1/2,0)}(\TR) = \sigma_{1}(\TR) + \sigma_{2}(\TR)\,,
\end{gather}
with
\begin{align}
\sigma_{1}(\TR) &:= \sum_{n=0}^{\infty} a_{\ell,n}\,\frac{1}{\TR^{\ell+1+n}}\,,
\\
\sigma_{2}(\TR) &:= \sum_{n=0}^{\infty} b_{\ell,n}\,\frac{1}{\TR^{\ell+4+2n}}\,,
\end{align}
into Eq.~\eqref{ScalarEOM3} and find recursion relations for the
$a_{\ell,n}$ and $b_{\ell,n}$ coefficients.

The recursion relations for the $a_{\ell,n}$ can be solved to obtain
\begin{gather}
  a_{\ell,n} = \frac{(\ell+n)!}{\ell!\,n!}\,a_{\ell,0}\,,
\end{gather}
which then leads to
\begin{gather}\label{eq:LeadingTerm}
  \sigma_{1}(\TR) = \frac{a_{\ell,0}}{(\TR-1)^{\ell+1}} ~.
\end{gather}
Since this is the leading behavior of the scalar field at large $\TR$,
we can determine the coefficient $a_{\ell,0}$ by comparing it with the
incomplete Beta function representation of the previous subsection:
\bw
\begin{multline}
  a_{\ell,0} = - \sum_{n=0}^{\infty} \alpha_{\ell,n} \, B(2n+3,\ell+1)
   = \frac{(-1)^{\tfrac{\ell+1}{2}} \sqrt{\pi}\,\ell!}{2^{\ell}}\,\left[\frac{(\ell+2)}{\Gamma(\ell+\tfrac{3}{2})}\,{}_{2}F_{1}\big(\tfrac{3}{2},2;\ell+\tfrac{3}{2};-1\big) \right. \\ \left. - \frac{6}{\Gamma(\ell+\tfrac{5}{2})}\,{}_{2}F_{1}\big(\tfrac{5}{2},3;\ell+\tfrac{5}{2};-1\big)\right]\,.
   \label{eq:aL0}
\end{multline}

Resumming the coefficients $b_{\ell,n}$ is more complicated. We can
solve the recursion relations to express the $b_{\ell,n}$ coefficient
as finite sums that depend on the coefficients $\alpha_{\ell,n}$ in
the series expansion of the source:
\begin{align}
  b_{\ell,n} &= \frac{\Gamma(\ell+4+n)}{\Gamma(4+n)}\,\sum_{j=0}^{j_{max}} \frac{\Gamma(3+2j)}{\Gamma(\ell+4+2j)}\,\alpha_{\ell,j} - \frac{\Gamma(\ell+4+n)}{\Gamma(2\ell+5+n)}\,\sum_{j=0}^{j_{max}} \frac{\Gamma(2\ell+4+2j)}{\Gamma(\ell+4+2j)}\,\alpha_{\ell,j} ~,
  \label{eq:bLn}
\end{align}
where $j_{max} = n/2$ if $n$ is even, and $j_{max}=(n+1)/2$ if $n$ is
odd.

When one tries to perform the full infinite sum of the $b_{\ell,n}$
coefficients over $n$ to find $\sigma_{2}(\TR)$, one finds a familiar
problem: the coefficients of Eq.~\eqref{eq:bLn} are finite sums with
an upper limit that depends on $n$, which must then be summed to
infinity. To get around this problem, we can rewrite each finite sum
as the difference of two infinite sums:
\begin{gather}
  b_{\ell,n} = \frac{\Gamma(\ell+4+n)}{\Gamma(4+n)}\,c_{\ell,j_{max}} - \frac{\Gamma(\ell+4+n)}{\Gamma(2\ell+5+n)}\,d_{\ell,j_{max}} ~,
\end{gather}
where
\begin{align}
  c_{\ell,k} ={}& (-1)^{\tfrac{\ell-1}{2}} \frac{\sqrt{\pi}}{2^{\ell+1}} \times \left[ \frac{(\ell+1)(4\ell-7)}{2\,\Gamma(\ell+\tfrac{3}{2})} + \frac{2\,(-1)^{k} (2k+3) \Gamma(k+\tfrac{5}{2})}{\sqrt{\pi}\,\Gamma(\ell+\tfrac{3}{2}+k)} \right. \\ &{}+ \frac{(\ell^4 - 2\ell^2 + 9\ell+10)}{2\,\Gamma(\ell+\tfrac{5}{2})}\,{}_{2}F_{1}(-\tfrac{1}{2},1;\ell+\tfrac{5}{2};-1) 
  - \frac{(\ell^4 + 5\ell^3 + \ell+5)}{2\,\Gamma(\ell+\tfrac{5}{2})} \,{}_{2}F_{1}(\tfrac{1}{2},1;\ell+\tfrac{5}{2};-1) \nn\\ 
& {} + \frac{2\,(-1)^{k}\,(\ell-1)\,\Gamma(k+\tfrac{5}{2})}{\sqrt{\pi}\,\Gamma(\ell+\tfrac{5}{2}+k)} \,{}_{2}F_{1}(1,k+\tfrac{5}{2};\ell+k+\tfrac{5}{2};-1)   \left. + \frac{2^{\ell-1} (\ell+2)}{\Gamma(\ell+\tfrac{3}{2})} \,{}_{2}F_{1}(\ell-\tfrac{1}{2},\ell;\ell+\tfrac{3}{2};-1) \right]\,,\nn\\
d_{\ell,k} ={}& (-1)^{\tfrac{\ell-1}{2}}\times  \left[ -\frac{1}{2}\,\Gamma(\ell+3) + \frac{(-1)^{k} 2^{\ell+2} (k+1) (\ell+k+2) \Gamma(\ell+k+3)}{\Gamma(k+2)}\right. \\ \nonumber
& \left. \quad + \frac{(-1)^{k} 2^{\ell+1} (\ell+2) \Gamma(\ell+k+3)}{\Gamma(k+2)}\,{}_{2}F_{1}(1,\ell+k+3;k+2;-1) \right]
\,.
\end{align}
With the $b_{\ell,n}$ coefficients expressed in this form, the second
sum for the scalar field becomes
\begin{align}
\sigma_{2}(\TR) &= \sum_{n=0}^{\infty} \left[ \frac{\Gamma(\ell+4+n)}{\Gamma(4+n)}\,c_{\ell,j_{max}} - \frac{\Gamma(\ell+4+n)}{\Gamma(2\ell+5+n)}\,d_{\ell,j_{max}} \right] \frac{1}{\TR^{\ell+4+2n}}\,.
\end{align}
We have not succeeded in finding closed-form expressions for the sum
over $n$ given a generic $\ell$ value, but the sum can be performed
for a given value of $\ell$.  \ew

\section{Series Solutions for the Trace of the Metric Perturbation}
Instead of truncating the Legendre expansion of the scalar field, it
is also possible to construct series approximations of the modes
$\tg_{\ell}^{\SOZO}$. We first note that the source term
\eqref{eq:SLdefinition} can be expanded in powers of $1/\TR$. For the
$\ell=0$ mode this series takes the form
\begin{gather}\label{eq:L0SourceSeriesExp}
  S_{0}^{\SOZO}(\TR) = \sum_{n=0}^{\infty} e_{0,n}\,\frac{1}{\TR^{4+n}} ~,
\end{gather}
while for $\ell \geq 2$ it is
\begin{gather}
  S_{\ell}^{\SOZO}(\TR) = \sum_{n=0}^{\infty} e_{\ell,n}\,\frac{1}{\TR^{\ell+2+n}}.
\end{gather}
In terms of the series coefficients for the source, the $\ell=0$ mode is
\begin{equation}
  \tg_{0}^{\SOZO}(\TR) = \sum_{n=0}^{\infty} e_{0,n} \times \Bigg[ \log \left(\frac{\TR-1}{\TR}\right) + \sum_{j=1}^{n+3} \frac{1}{j\,\TR^{j}} + \frac{1}{n+3}\,\frac{1}{\TR-1}\,\left(\frac{1}{\TR^{n+3}} - 1\right)\Bigg] ~.
\end{equation}
Note that the coefficient of the $\log(\TR-1)$ term is
\eqref{eq:L0SourceSeriesExp} evaluated at $\TR=1$, as in
Eq.~\eqref{eq:MonopoleModeAtHorizon}. The modes with $\ell \ge 2$ can
be expressed as a series involving incomplete Beta functions:
\begin{multline}
  \tg_{\ell}^{\SOZO}(\TR) = \sum_{n=0}^{\infty} e_{\ell,n} \times \Big[ - (\TR-1)^{\ell}\,B_{1/\TR}\left(2\ell+2+n,-\ell\right) \\
  {}- \frac{1}{(\TR-1)^{\ell+1}}\,B_{1-1/\TR}\left(\ell+1, n+1\right) \Big] ~.
\end{multline}
Since these solutions are obtained directly from
Eq.~\eqref{eq:g2-FirstSolution} they already satisfy the correct
boundary conditions at $\TR \to \infty$ and $\TR =1$.

Given the expansion of the source functions, one can obtain a series
solution of the equation of motion Eq.~\eqref{eq:glModeEquation}
directly. For the $\ell=0$ mode this solution takes the form
\begin{equation}
  \tg_{0}^{\SOZO}(\TR)= \frac{f_{0,0}}{(\TR-1)} + \sum_{n=0}^{\infty} \frac{1}{\TR^{4+n}} \sum_{j=0}^{n} \frac{n+1-j}{(n+4)(j+3)}\,e_{0,j}
\end{equation}
while for $\ell \geq 2$ it is
\begin{multline}
  \tg_{\ell}^{\SOZO}(\TR) = \frac{f_{\ell,0}}{(\TR-1)^{\ell+1}} + \sum_{n=0}^{\infty} \frac{(\ell+n)!}{\TR^{\ell+2+n}}
  \sum_{j=0}^{n} e_{\ell,j}\,\left(\frac{j!}{(n+1)!(\ell+j+1)!} \right. \\  - \left. \frac{(2\ell+j+1)!}{(\ell+j+1)!(2\ell+n+2)!}\right) ~.
\end{multline}
In both cases the coefficient $f_{\ell,0}$ of the leading term can be
expressed in terms of one or more integrals of the source; for
$\ell = 0$ it is
\begin{gather}
  f_{0,0} = \int_{\infty}^{1} d\TR\, S_{0}^{\SOZO}(\TR) ~.
\end{gather}

\bibliographystyle{iopart-num}
\bibliography{master}

\end{document}